\documentclass[sigconf]{acmart}

\usepackage{amsmath}
\usepackage{listings}
\usepackage{xspace}
\usepackage{bm}
\usepackage{subfig}
\usepackage{wrapfig}
\usepackage{graphicx}
\usepackage{comment}
\usepackage{multirow}
\usepackage{booktabs}
\usepackage{mathtools}
\usepackage{amsthm}
\usepackage{bm}
\usepackage{xcolor}
\usepackage{subcaption}
\usepackage[leftcaption]{sidecap}
\usepackage{xurl}
\usepackage{algorithm}
\usepackage{algpseudocode}
\usepackage{balance}

\AtBeginDocument{%
  }

\newcommand\sys{$\mathsf{Sentry}$\xspace}

\newcommand{\Paragraph}[1]{\vspace{.3em}\noindent\textbf{#1}}

\definecolor{codegreen}{rgb}{0,0.6,0}
\definecolor{codegray}{rgb}{0.5,0.5,0.5}
\definecolor{codepurple}{rgb}{0.58,0,0.82}
\definecolor{backcolour}{rgb}{0.95,0.95,0.92}

\lstdefinestyle{mystyle}{
    backgroundcolor=\color{backcolour},   
    commentstyle=\color{codegreen},
    keywordstyle=\color{magenta},
    numberstyle=\tiny\color{codegray},
    stringstyle=\color{codepurple},
    basicstyle=\ttfamily\footnotesize,
    breakatwhitespace=false,         
    breaklines=true,                 
    frame=ltb,
    framerule=0pt
}

\copyrightyear{2025}
\acmYear{2025}
\setcopyright{cc}
\setcctype{by}
\acmConference[CCS '25]{Proceedings of the 2025 ACM SIGSAC Conference on Computer and Communications Security}{October 13--17, 2025}{Taipei, Taiwan}
\acmBooktitle{Proceedings of the 2025 ACM SIGSAC Conference on Computer and Communications Security (CCS '25), October 13--17, 2025, Taipei, Taiwan}\acmDOI{10.1145/3719027.3765070}
\acmISBN{979-8-4007-1525-9/2025/10}

\begin{document}


\title{Sentry: Authenticating Machine Learning Artifacts on the Fly}
%
 \author{Andrew Gan}
 \email{gan35@purdue.edu}
 \affiliation{%
   \institution{Purdue University}
   \city{West Lafayette}
   \state{IN}
   \country{USA}
 }

 \author{Zahra Ghodsi}
 \email{zahra@purdue.edu}
 \affiliation{%
   \institution{Purdue University}
   \city{West Lafayette}
   \state{IN}
   \country{USA}
 }


\begin{abstract}
Machine learning systems increasingly rely on open-source artifacts such as datasets and models that are created or hosted by other parties. The reliance on external datasets and pre-trained models exposes the system to supply chain attacks where an artifact can be poisoned before it is delivered to the end-user. Such attacks are possible due to the lack of any authenticity verification in existing machine learning systems. Incorporating cryptographic solutions such as hashing and signing can mitigate the risk of supply chain attacks. However, existing frameworks for integrity verification based on cryptographic techniques can incur significant overhead when applied to state-of-the-art machine learning artifacts due to their scale, and are not compatible with GPU platforms.
In this paper, we develop \sys, a novel GPU-based framework that verifies the authenticity of machine learning artifacts by implementing cryptographic signing and verification for datasets and models. \sys ties developer identities to signatures and performs authentication on the fly as artifacts are loaded on GPU memory, making it compatible with GPU data movement solutions such as NVIDIA GPUDirect that bypass the CPU. \sys incorporates GPU acceleration of cryptographic hash constructions such as Merkle tree and lattice hashing, implementing memory optimizations and resource partitioning schemes for a high throughput performance. Our evaluations show that \sys is a practical solution to bring authenticity to machine learning systems, achieving orders of magnitude speedup over a CPU-based baseline.
\end{abstract}




\maketitle

\section{Introduction}
Machine learning (ML) supply chains describe the complex series of steps carried out by multiple parties which characterize dataset generation, training, fine-tuning, testing, and distribution of ML products. Similar to software supply chains, ensuring the security of ML supply chains is critical to the overall security of an ML product. Unfortunately, unlike software supply chain security which has been the subject of much prior work~\cite{geer_quant_2020,newman2022sigstore,torres2019toto}, ML supply chain security has only recently gained attention. 
Recent work has identified ML supply chain attacks that can compromise the security of ML systems.
Carlini et al.~\cite{carlini_dataset_2024} discovered that datasets published online and hosted through a URL can be poisoned when an adversary obtains an expired domain name and gains the ability to modify the hosted data arbitrarily.
Gu et al.~\cite{gu_badnets_2017} introduced backdoored neural networks which behave maliciously on inputs with an attacker chosen backdoor trigger, and can be covertly substituted with benign models in ML supply chains. Similarly, maliciously modifying a pre-trained model can enable an adversary to attack users fine-tuning the model for specific applications and fully compromise the privacy of the fine-tuning data~\cite{feng_backdoors_2024}. 
Similar supply chain attacks have already occurred in the real world~\cite{grosse2024your} with a recent example of an insider attack where an intern sabotaged internal ML models at ByteDance~\cite{bytedance}.
In the examples above, the attacks are possible due to a lack of authenticity verification of the artifacts (i.e., datasets or models) in existing ML systems.

A large body of work exists to secure generic software supply chains~\cite{reichert2024software,torres2019toto,newman2022sigstore}, allowing developers to verify certificates of operations and artifacts across the supply chain steps.
In practice, the unique characteristics of ML supply chains including involved artifacts (datasets and models) and backend computing platforms (GPU servers) introduces challenges in applying software supply chain security solutions to ML supply chains. For example, when using Sigstore~\cite{newman2022sigstore} to sign a model artifact, the model files should be serialized into a stream of bytes and passed sequentially through a cryptographic hash function to obtain a short digest which is then incorporated in a payload and cryptographically signed. This approach has several drawbacks. 
The feasibility of insider attacks demonstrates the need for verifying the integrity of artifacts regularly and ideally every time before use. However, sequential CPU hashing can incur significant overhead when applied to state-of-the-art ML models with millions or billions of parameters~\cite{narayanan2021efficient}. For example, the GPT2-XL has over 1.6 billion parameters and takes 34  seconds to be saved as a PyTorch pth file and up to 26 seconds to hash depending on the hashing algorithm. This incurred overhead negatively impacts the throughput of ML systems and can create scenarios where users have to choose between efficiency and security. Additionally, using CPU-based solutions requires the ML artifacts (e.g., the model) to be loaded into CPU memory to be hashed. This requirement is not compatible with efficient GPU data movement solutions such as NVIDIA GPUDirect~\cite{gpudirect} which bypasses the CPU and can load models from storage or remote GPUs directly into GPU memory. 
As such, a solution to authenticate the integrity of ML artifacts on the GPU is required.

In this paper, we present \sys, the first framework to authenticate the integrity of machine learning artifacts on the fly when they are loaded into GPU memory. 
\sys closes up the gaps in securing ML supply chains by providing an efficient solution to artifact authentication which is scalable to state-of-the-art models and datasets.
\sys uses the Compute Unified Device Architecture (CUDA) to implement novel GPU architectures for cryptographic hash constructions, incorporating dedicated memory optimizations and new adaptive resource partitioning schemes for high throughput performance. \sys proposes new design strategies to enable granular integrity checks for ML artifacts (such as per-layer or per-model). We instantiate \sys in Python for ease of use by ML practitioners
 as a generic framework for ML artifact trust on the GPU for ML supply chain security solutions.


In summary, this paper makes the following contributions:

\begin{itemize}
    \item GPU implementation of cryptographic hash constructions based on \emph{Merkle tree} and \emph{lattice hashing} supporting various underlying hash functions and incorporating GPU optimization strategies such as shared memory computation and streaming for pipelining operations. 

    \item Model and data authentication modules implemented on GPU and specifically tailored to ML systems with memory fragmentation compatibility and support for mixed-source datasets. Our on-the-fly artifact authentication achieves up to 269$\times$ and 180$\times$ speedup compared to a CPU-based baseline for model and datasets respectively.
    
    \item A Python-based and modular library, namely \sys, for \emph{ML artifact authentication on GPU} with support for efficient GPU data movement solutions such as GPUDirect for both model loading and data processing pipeline.
    
\end{itemize}

\section{Background}

The machine learning supply chain (MLSC) describes the operations from dataset collection and preparation, to model training, model evaluation, and service deployment. An MLSC can be described by \emph{artifacts} including datasets, algorithms, models and configurations, and ML frameworks which are used during its life cycle. Due to the scale and complexity of curating datasets and training models, ML hubs (such as ONNX Model Zoo~\cite{onnxmodels}, Hugging Face~\cite{hfmodels}, Kaggle~\cite{kmodels}, ModelScope~\cite{modelscope}, etc.) which store and distribute ML artifacts have gained popularity. These ML hubs provide inputs to MLSCs at different steps or host products generated from them.

ML artifacts can be attacked at different stages leading to system security compromises which necessitate integrity verification at every step of the pipeline. In this work, we focus on authentication of datasets and models which we refer to as ``ML artifacts''.

\subsection{Software Artifact Signing} \label{sec:swsign}
Software artifact signing has emerged as a promising mitigation for software supply chain attacks. A signature over a software package or artifact allows a user to verify the integrity of the artifact they receive and ensure that it has not been tampered with en route.
Sigstore~\cite{newman2022sigstore} is a framework for signing software to achieve the aforementioned goal. Sigstore uses OpenID Connect (OIDC) with a public-key infrastructure (PKI) for user authentication, incorporates digital signatures for artifacts, and includes logging infrastructure (i.e., transparency log) for identities and artifact signatures which are accessed during verification process. During signing and verification, information regarding how an artifact is signed is inserted into an attestation payload.
\begin{wrapfigure}{r}{.2\textwidth}
\centering
\includegraphics[width=\linewidth]{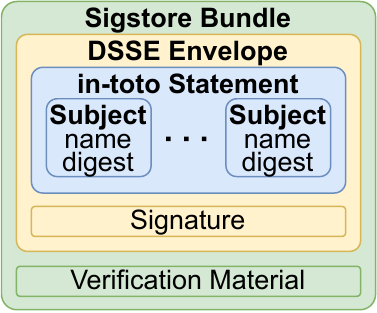}
\caption{Sigstore payload structure.}
\label{fig:payload}
\end{wrapfigure}
Figure~\ref{fig:payload} illustrates the structure of Sigstore attestation payload that we adopt in this work. The payload is a JSON formatted data structure where the innermost layer is an in-toto Statement~\cite{torres2019toto,intotospec} containing Subject arrays with information about the artifacts the attestation applies to and their digest values. The next layer wraps the in-toto Statement with the signature using a DSSE Envelope~\cite{dsse}. The outermost layer constitutes the Sigstore Bundle including signature verification material such as certificate chains or a public key together with the DSSE Envelope.
Combined with the transparency log, the attestation payload enables secure verification by any party who wishes to use the signed software and ensure its authenticity. 
To sign an artifact, Sigstore computes the hash of the (serialized) content to obtain a short digest, which it then signs (for example using the ECDSA~\cite{johnson2001ecdsa} algorithm). The end-user obtaining the artifact repeats the same steps of computing the content hash and verifies the signature. To accommodate the scale of ML artifacts, we focus on artifact hash computation and provide a brief primer on cryptographic hashing algorithms next.

\subsection{Cryptographic Hash Constructions} \label{sec:HashConstr}
\begin{figure}
    \centering
    \includegraphics[width=\linewidth]{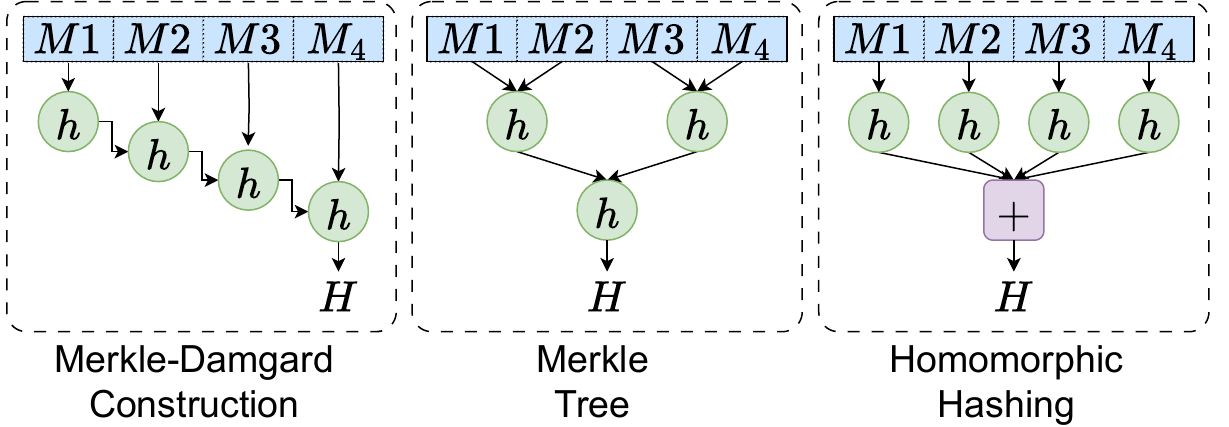}
    \caption{Cryptographic hash constructions based on Merkle-Damg{\aa}rd, Merkle tree, and homomorphic hashing from one-way compression function $h$.}
    \label{fig:chf}
\end{figure}

Cryptographic hash functions (CHF) map arbitrary length data to a short fixed-length digest which satisfy specific security properties such as collision resistance (i.e., it is difficult to find two inputs that produce the same output) and preimage resistance (i.e., it is difficult to find an input for a given output)~\cite{rogaway2004chf}. CHFs are constructed from \emph{one-way compression functions} which map a fixed-length input to a fixed-length digest. Compression functions are designed to be easy to compute but hard to reverse. Examples of compression functions include SHA-2~\cite{nist_sha256}, SHA-3~\cite{cryptoeprint:2015/389}, and Blake2~\cite{cryptoeprint:2013/322}. Several construction exist which create a CHF from a compression function as described below and summarized in Figure~\ref{fig:chf}.

\Paragraph{Merkle–Damg{\aa}rd:} In Merkle-Damg{\aa}rd construction~\cite{merkle1989one,damgaard1989design}, a padding function is applied to the message to be hashed in order to create an input with size which is a multiple of a fixed number, namely the block size. The compression function is then applied sequentially to each data block, each time combining a block of input with the output from the previous round (set as an initial fixed value for the first block). 
The final hash output is the message digest.

\Paragraph{Merkle Tree:} The Merkle tree~\cite{merkle1979secrecy} is a tree-based hashing algorithm which starts with dividing the input message into blocks and computing the digest of each block independently using the compression function. The resulting digests are then reduced to one in a tree structure. At each tree level, two neighboring nodes are processed by the compression function to obtain one output node.  If a layer has an odd number of nodes, a padding node is added to the layer. Finally, the root node is the message digest. The Merkle tree construction provides several benefits. First, it eliminates the sequential dependency of the digest computation in Merkle-Damg{\aa}rd and is more suitable for parallelization. Second, if a message block is updated, only the path from that block to the root node has to be redone instead of the entire process. This is specially important for database applications where some entries or data points may be updated at a time, yielding a more efficient final digest update computation.

\Paragraph{Homomorphic Hashing:} Homomorphic hashing constructions~\cite{bellare1994incremental} extend the idea of efficient digest update when a small change in the input message occurs. Bellare and Micciancio~\cite{bellare1997new} proposed specific constructions under this paradigm based on the hardness of specific computational problems in groups. One of their proposed constructions was the \emph{lattice hash} which was later formalized in subsequent work~\cite{lewi2019securing}. Lattice hash applies a compression function to message blocks, and combines the resulting digests with a component-wise vector modular addition. An important property in lattice hash is set homomorphism, meaning that for two disjoint sets $S_1$ and $S_2$ we have $\texttt{LtHash}(S_1) + \texttt{LtHash}(S_2) = \texttt{LtHash}(S_1 \cup S_2)$. It is also true that $\texttt{LtHash}(S_1 \cup S_2) - \texttt{LtHash}(S_1) = \texttt{LtHash}(S_2)$. This set homomorphism property results in very efficient digest update computation. Additionally, similar to the Merkle tree construction, lattice hash computation is highly parallelizable. The lattice addition of $k$ inputs where $M_{i}$ is the $i$-th input, with each input broken into $n$ partitions of $d$ bits, can be expressed as:
\begin{equation}
\texttt{LtHash}_{n,d}(\{M_{1},...,M_{k}\}) = \sum_{i=1}^{k}{h(M_{i})}\mod{q}    
\end{equation}
where $q$ is the modulo and $h$ is the compression function. Lewi et al.~\cite{lewi2019securing} set $q=2^d$ which results in a simple modular operation by discarding any carry-over values after addition. 

The addition operation in lattice hash is \emph{order invariant}, which means that the order in which individual digests are accumulated into the final value does not change the result. Therefore, message block indices are incorporated into the digests before addition for security. Nevertheless, the order invariant property of lattice hash provides optimization opportunities on GPU which we will detail later.



\subsection{GPU Architecture} \label{sec:GPUArchitecture}
We provide a brief summary of the main GPU features, and refer the readers to other references for more detail~\cite{cudaguide}. 
\Paragraph{GPU Execution Model:} GPUs follow a single instruction multiple thread execution model, enabling massive thread-level parallelism. Tasks offloaded to the GPU are initiated by the CPU through asynchronous kernel calls. Any data dependency in a sequential step then requires synchronization with the CPU.
In NVIDIA GPUs, threads are organized hierarchically into warps, blocks, and grids. Warps are the smallest execution unit consisting of 32 threads, and should access coalesced memory with minimal control divergence for optimal performance. A thread block consists of up to 1024 threads assigned to a streaming multiprocessor. Multiple blocks are combined to form a grid. 
A program performance can be improved through the use of \emph{CUDA streams}. A stream is a queue of GPU operations which are executed in a specific order. Different streams can be executed concurrently, allowing for overlapping of memory transfers and kernel executions for an improved performance.


\Paragraph{GPU Memory:} A GPU is composed of a memory hierarchy with different access times and capacities. Those are the registers, shared memory, global memory, and constant memory. Registers are local to each thread and quickest to access. Shared memory is shared between threads in a block and are slower to access than registers. Global memory is the largest memory and stores data for the lifetime of the program. It has the longest access time. Constant memory is a read-only memory which allows multiple threads to read from the same address through broadcasting its value to all reading threads at once.

\Paragraph{GPUDirect:} To enhance data movement for GPUs and mitigate the increasing time spent loading data, NVIDIA has developed a family of technologies called GPUDirect~\cite{gpudirect}. This technology provides a direct path between storage and GPU memory (GPUDirect Storage) as well as between GPUs and a third party device through Remote Direct Memory Access (GPUDirect RDMA). Therefore, data movement can bypass the CPU and eliminate required buffer copies of data, improving performance.


\begin{figure*}[ht!]
  \includegraphics[width=\textwidth]{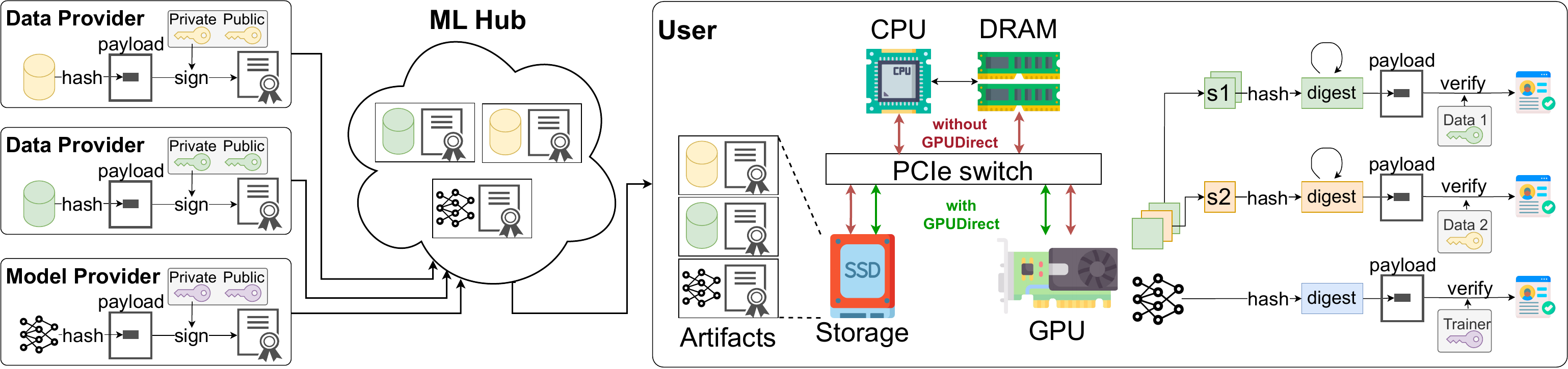}
  \caption{High-level description \sys with data provider, model provider, ML hub, and user roles. Model and data providers augment their artifact with an authenticated payload before uploading them to the hub. Any user accessing artifacts on the hub can verify their authentication on the fly as artifacts are loaded on GPU memory for ML tasks.}
  \label{fig:e2esystem}
\end{figure*}

\section{System and Threat Model}
In this section, we detail our setting including the parties involved and describe attacker capabilities and goals.

\subsection{System Model}
We consider the following roles:

\begin{itemize}
    \item \textbf{ML Hubs} are systems that host and distribute ML artifacts such as models and datasets.
    \item \textbf{Dataset Providers} are parties collecting or preparing datasets for ML tasks (training or inference) which they can upload to ML hubs.
    \item \textbf{Model Providers} are parties training models which they upload as pre-trained models to ML hubs.
    \item \textbf{Users} are parties that obtain an ML artifact from a hub (dataset or model) and use it for ML tasks (inference or fine-tuning).
\end{itemize}

We simplify the system description by abstracting out the parties that provide identity or monitoring services such as certificate authorities, identity and artifact logs, and log monitors as described in Section~\ref{sec:swsign}, and refer the readers to corresponding systems~\cite{newman2022sigstore} for more detail.

\subsection{Threat Model}

The adversary's goal is to implant a tampered ML artifact into the supply chain, such as backdoored models~\cite{gu2019badnets} or poisoned datasets~\cite{carlini2024poisoning} which can be carried out by internal~\cite{bytedance} or external~\cite{gu2019badnets} attackers.
We consider a scenario where the adversary tampers with the ML artifacts at any point before they are loaded into GPU memory, but assume secure \sys users and providers running signing and verification as well as existence of a secure public key infrastructure. We envision the following types of compromise:
\begin{enumerate}
    \item An adversary can perform man-in-the-middle attack targeting an artifact in transit between an artifact provider and a hub, or between a hub and the user.
    \item An adversary can tamper with the artifacts residing on user storage.
    \item A hub can be compromised by an adversary.
\end{enumerate}
    
ML artifact compromises can sabotage the downstream ML system performance and undermine security guarantees. \sys lays the groundwork for a paradigm where authentication is always performed at runtime which is practical due to \sys’s small overhead.
We note that the assumptions on the secure user can be further relaxed using GPU-based secure enclaves~\cite{volos2018graviton,yudha2022lite} as we discuss in Section~\ref{sec:related}.

\section{\sys Framework}
\subsection{Overview}
In this section, we provide a high-level overview of \sys design as depicted in Figure~\ref{fig:e2esystem}. The signing ecosystem in \sys is implemented through Sigstore~\cite{newman2022sigstore} which provides mechanisms to bind signatures to identities and incorporates signing procedure attestation as described in Section~\ref{sec:swsign}.
Accordingly, \sys creates authenticated payloads for each ML artifact.
\sys consists of two main components for model and dataset authentication. We describe the signing and verification flow for each component below.\\

\subsubsection{Model Authentication}\label{sec:modelauth}
As previously mentioned, existing solutions for software signing require the artifact to be loaded into CPU memory for hash and signature computation which incurs significant overhead for state-of-the-art ML systems as we demonstrate later in Section~\ref{sec:modelruntime}. \sys overcomes this limitation by performing authentication operations on ML artifacts directly on the GPU. 

We start by describing the signing flow for a model provider, assuming model training is completed or model provider is saving a checkpoint of the model currently residing on GPU memory.
\sys generates the attestation payload associated with the model on the CPU with dummy digests. Then, \sys moves the payload into GPU memory, computes model digests on GPU, and inserts the  digests into the payload. The private key associated with the signing key-pair is also copied to GPU memory. Afterwards, \sys hashes the payload and signs the payload digest on GPU using the key, and obtains the signature.

\sys implements several hashing constructions and provides model digest computation at two granularity levels. The first granularity level only computes a single digest for the entire model. The second granularity level computes per-layer digest computation in addition to the final digest which will be included in the authenticated payload. This flexibility enables \sys to be applicable in settings where an audit trail is required to verify computations such as fine-tuning where only a subset of model layers are modified and the rest are frozen. 
In order to verify a pre-trained model obtained from a third party or a hub, a user applies \sys to recompute model digests on the GPU and verify the signature on the associated payload. \sys incorporates several GPU optimizations by performing efficient memory accesses when hashing an ML model loaded in fragmented GPU memory, in addition to carefully designing concurrent computations for a highly efficient runtime. 
In addition to runtime benefits, \sys can authenticate models within ML systems that integrate efficient GPU data movement solutions such as GPUDirect Storage which bypass the CPU as shown in Figure~\ref{fig:e2esystem}.

\subsubsection{Dataset Authentication}\label{sec:dataauth}
Dataset authentication in ML systems faces additional challenges since datasets can originate from multiple sources and used within the same ML pipeline. For example, a training dataset can be aggregated by combining several datasets, or a server might perform inference computation on batched queries from several clients. Furthermore, training datasets (e.g., images) are often compressed before they are stored, requiring developers to build multi-stage data processing pipelines that load, decode, and perform transformations such as cropping or resizing on datasets. These data processing pipelines, typically executed on the CPU, have become a bottleneck. Including an authentication step into the data processing pipeline will exacerbate this problem and limit throughput further. Finally, ML training pipelines typically include a random shuffling of the data which in the case of dataset sourced from multiple data providers, means that each data batch loaded to the GPU could contain data points from multiple providers.

\sys is designed to address the above challenges, providing the flexibility required to preserve the functionality of ML operations in addition to authentication guarantees with minimal overhead.
For dataset authentication, \sys augments each data point to include a digest in addition its value (raw data), label, and provider source ID. The digest is computed on raw data prior to any encoding step for storage. For example, the digest of an image is the hash computed over the 3D byte tensor. Each data provider will then generate an attestation which contains the final digest over the entire dataset and is signed by the data provider. \sys incorporates lattice hashing for dataset digest computation which enables simple digest updates when individual data points are added or removed. Utilizing lattice hashing for data authentication will further enable \sys to authenticate randomly shuffled datasets efficiently as detailed next.

To enable efficient data processing and mitigate the added overhead of data authentication, \sys integrates NVIDIA Data Loading Library (DALI)~\cite{nvdali} which accelerates data processing operations including loading, decoding, and transformations by offloading them to the GPU, resulting in improved performance and scalability for training and inference. Within DALI, data authentication is performed as an additional step in the data processing pipeline as depicted in Figure~\ref{fig:e2esystem}. Specifically, \sys keeps track of a running sum of digests for each data source using lattice hash as batches of data are loaded into the GPU. For each batch, \sys uses a dictionary to map each data source to its data points in the batch, and computes per-source digests before adding them to their corresponding digest running sum buffers. Afterwards, the data batch goes through the rest of the pipeline which implements the remaining data processing operations and performs a training or inference round.
Once all batches from the dataset are processed, the digest buffers contain the final dataset digests and are used in signature verification.



In the next section, we detail the implementation of \sys digest computation on GPU and discuss incorporated design optimizations. 
We start by describing algorithms for hashing specifically tailored to ML model artifacts, followed by \sys data authenticator component.

\begin{figure}
\centering
\includegraphics[width=\linewidth]{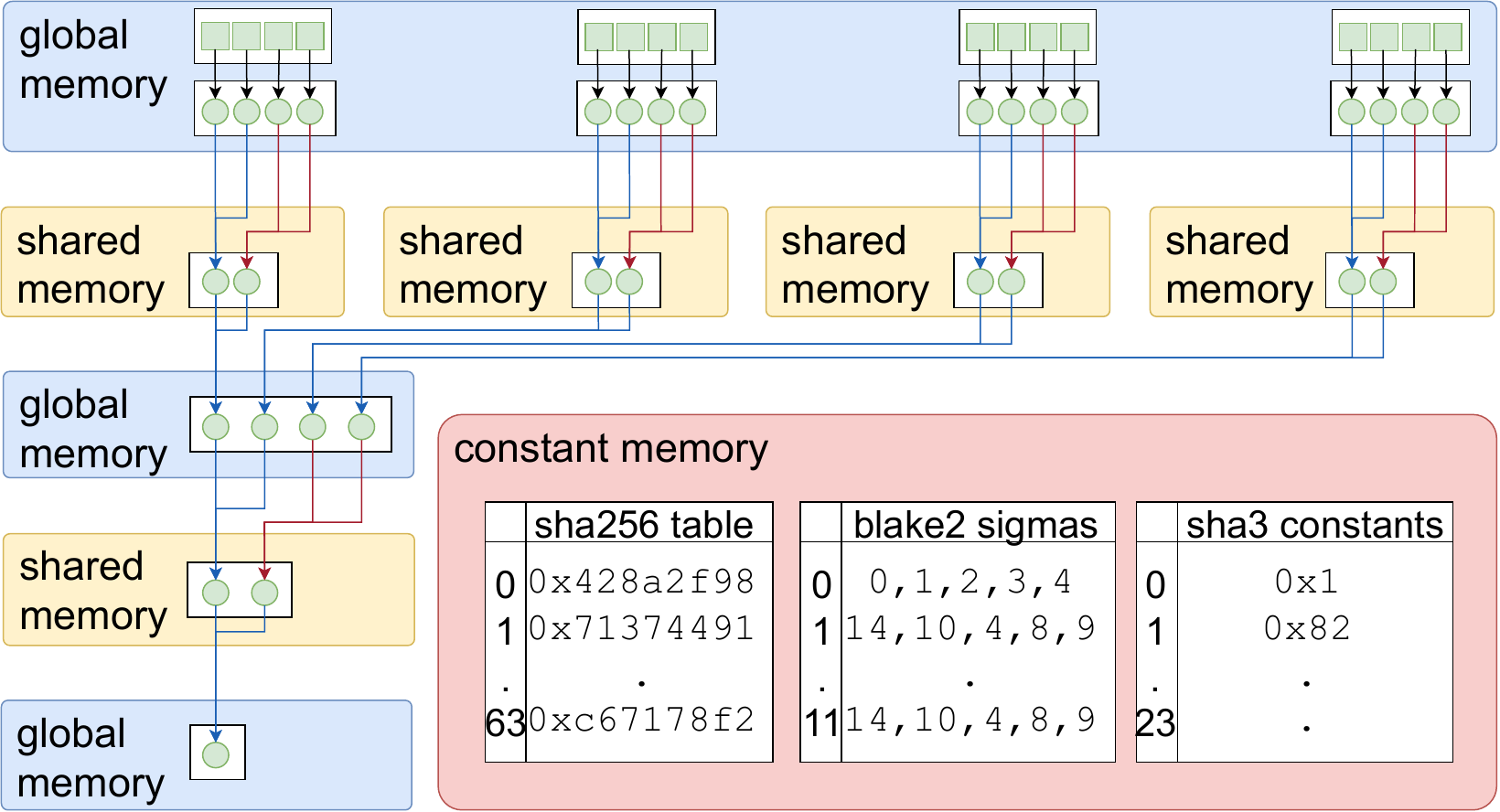}
\caption{Depiction of block hashing and tree reduction kernels. Shared memory is used for storing intermediate values within a thread block, global memory for collecting results from different blocks, and constant memory for storing lookup tables with hard-coded values. 
The input read from global memory is reduced in shared memory, whereas the last intermediate results in shared memory are reduced to the final result in global memory to minimize the number of memory transfers.}
\label{fig:blockReduction}
\end{figure}

\subsection{Model Hash Computation on GPU}

Efficient hash computation of large ML model artifacts requires exploiting parallelization opportunities on the GPU. To do so, we focus on two hash constructions based on Merkle tree and lattice hash introduced in Section~\ref{sec:HashConstr} which are amenable to parallel execution.

We start by describing how ML models are defined and loaded into GPU memory. 
In PyTorch, models are described using a Python dictionary object called a \texttt{state\_dict} which maps each layer to its parameter tensor.
PyTorch implements dynamic memory allocation optimizations on the backend with CUDA which uses the smallest available free blocks for allocation requests or allocates a new block. As a consequence, when a model is loaded into GPU memory, each parameter tensor is stored in a contiguous block of memory, but different layers could be fragmented across the memory space. We note that this behavior is not unique to PyTorch, and is implemented in other ML frameworks as well. For example, in Tensorflow each model layer is allocated a Tensorflow tensor, resulting in the scattering of model weights across GPU memory. Based on this insight, we propose different designs for hashing ML models on the GPU with different memory and runtime trade-offs. We start by describing our implementation of the Merkle tree hash on GPU, followed by our proposed architecture for lattice hash.


\begin{figure*}[t]
\centering
    \includegraphics[width=\linewidth]{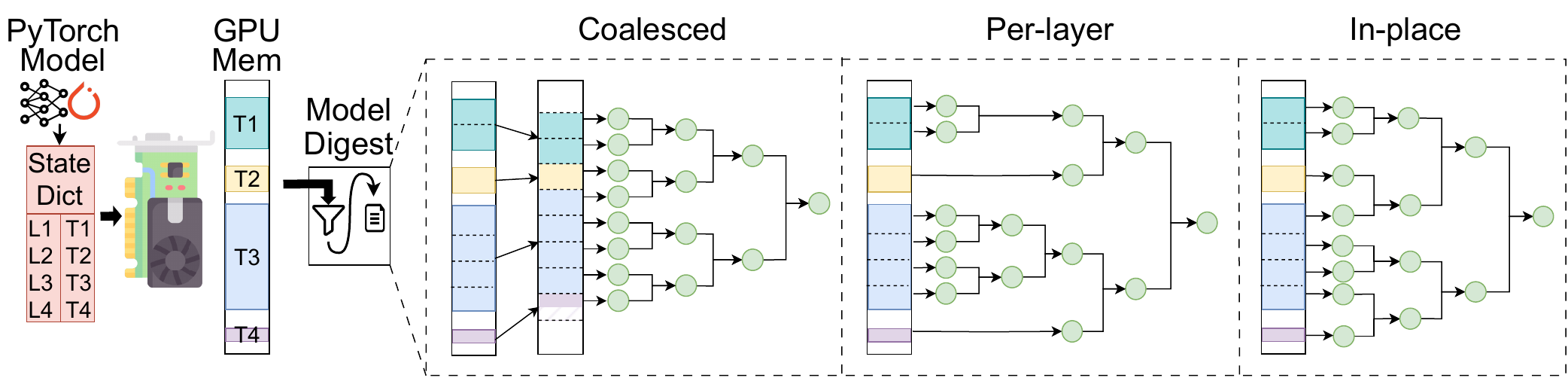}
    \caption{Depiction of three model hashing implementations. The model is defined in PyTorch using a dictionary \texttt{state\_dict}, which maps each layer to the corresponding weight tensor. After the model is loaded on GPU memory, the layers are fragmented across GPU memory space. We propose coalesced, per-layer, and in-place implementations for computing the model hash.}
    \label{fig:hashing}
\end{figure*}

\subsubsection{Merkle Tree Hashing on GPU} \label{sec:MerkleTree}
\sys implementation of the Merkle tree hashing consists of two efficient CUDA kernels as depicted in Figure~\ref{fig:blockReduction}. The first kernel, namely the \emph{block hashing kernel}, hashes fixed size data blocks in global memory into digests. In this kernel, each thread is assigned a single data block (with default size of 8192 bytes) to hash. The resulting digests are written to an output buffer in global memory.
The second kernel, namely the \emph{reduction kernel}, takes the output buffer of the hashing kernel as input, and in each step reduces the size in half through a Merkle tree layer in shared memory, finally generating the output digest as shown in Figure~\ref{fig:blockReduction}. We describe the details of the reduction kernel below.

The reduction kernel takes as input the digests to be reduced and outputs the reduced digests computed by each thread block. In other words, if a thread block has $k$ threads, each thread block reduces $2k$ digests into one digest. With $n$ digests as input to a reduction kernel, the output consists of $n/2k$ digests. The reduction kernel is repeatedly called until a single digest remains. To prevent unnecessary memory transfers, we use two temporary memory buffers and designate one as the input buffer and the other as the output buffer. At the end of each reduction kernel call, we switch the designation of buffers such that the output buffer (containing the layer digests) is designated as the input buffer and the previous input buffer is now the output buffer. As a result, the next call to the reduction kernel can start immediately without the need for any memory copy operations.

Within a thread block, each thread hashes two neighboring digests and stores the resulting digest into shared memory. As mentioned in Section~\ref{sec:GPUArchitecture}, shared memory allows for quicker access and resource sharing between threads in a thread block. Additionally, we use the constant memory to store the lookup tables for hardcoded values (such as the key values for SHA256 and sigma values for BLAKE2) used in hashing algorithms as it allows read-only access and can broadcast a value to multiple threads with one read. 
\sys supports SHA256, BLAKE2B, and SHA3 (based on~\cite{mochimodevCUDAHash}) as compression functions for Merkle tree hashing on the GPU.
In each reduction step, the number of threads within a thread block are reduced by half.
Afterwards, we perform an explicit synchronization of all threads in the thread block to prevent race conditions between tree layer reductions. Once the number of active threads falls below the warp size of 32, thread block synchronization is not needed, since threads in a warp execute instructions in lockstep, offering implicit synchronization. The final call to the reduction kernel generates the output digest of the entire Merkle tree.

Next, we detail hash computation of ML models based on our Merkle tree hashing kernel. As previously discussed, PyTorch optimizations for memory allocation results in model layers being fragmented across the memory space after being loaded on the GPU, making model digest computation and efficient memory accesses more challenging. We propose three architectures (namely coalesced, per-layer, and in-place) using Merkle hashing which address this issue as depicted in Figure~\ref{fig:hashing} and described next.

%

\begin{algorithm}[t]
\caption{\sys Coalesced Hashing}\label{alg:coalesced}
\begin{algorithmic}
    \State {\bfseries Input:} Model $\mathsf{state\_dict} = \{L_1:T_1,\dots,L_n:T_n\}$
    \State {\bfseries Output:} Model hash value
    \State $\mathsf{buffer} \leftarrow T_1\|\dots\|T_n$
    \State $m \leftarrow$ number of blocks in $\mathsf{buffer}$
    \For {$j=1$ in $m$}
        \State $\mathsf{hash\_blocks}[j]\leftarrow \mathsf{BlockHash}(\mathsf{buffer}[j])$
    \EndFor
    \State $\mathsf{hash} \leftarrow \mathsf{HashReduce}(\mathsf{hash\_blocks})$
    \State \textbf{return} \textsf{hash}
\end{algorithmic}
\end{algorithm}
\begin{algorithm}[t]
\caption{\sys Per-layer Hashing}\label{alg:perlayer}
\begin{algorithmic}
    \State {\bfseries Input:} Model $\mathsf{state\_dict} = \{L_1:T_1,\dots,L_n:T_n\}$
    \State {\bfseries Output:} Model hash value
    \For {$i=1$ to $n$}
        \State $m \leftarrow$ number of blocks in $T_i$
        \For {$j=1$ in $m$}
            \State $\mathsf{hash\_blocks}[i][j]\leftarrow \mathsf{BlockHash}(T_i[j])$
        \EndFor
        \State $\mathsf{layers\_hash}[i] \leftarrow \mathsf{HashReduce}(\mathsf{hash\_blocks[i]})$
    \EndFor
    \State $\mathsf{hash} \leftarrow \mathsf{HashReduce}(\mathsf{layers\_hash})$
    \State \textbf{return} \textsf{hash, layers\_hash}
 \end{algorithmic}
\end{algorithm}
 
\Paragraph{Coalesced Hashing}
The coalesced hashing scheme first coalesces the model layers into a contiguous block of memory by iterating through the model \texttt{state\_dict} and copying layer tensors into a contiguous memory block. The size of the contiguous memory block is a multiple of the data block size, and is padded at the end if the total model size is smaller. 
The details of the above steps for this hashing scheme are described in Algorithm~\ref{alg:coalesced}, where \textsf{buffer} represents the contiguous memory block.
Afterwards, the Merkle tree hashing kernel (described previously) operates on the contiguous memory block to obtain the final model digest. 
The coalesced scheme provides a baseline implementation and requires additional memory allocation to store the coalesced model, resulting in a 
2$\times$ memory requirement with respect to the model size. Furthermore, the memory copy operations to move layer tensors into the coalesced block impose additional runtime overhead. However, we note that GPU to GPU memory copy can be performed quite efficiently in modern GPUs which have a memory bandwidth above 700 GB/s.
%



\Paragraph{Per-layer Hashing}
The per-layer hashing scheme operates on each layer tensor (stored in a contiguous block of memory by PyTorch) separately to obtain per-layer digest values, which are then reduced into a single model digest. We outline the operations in per-layer hashing in Algorithm~\ref{alg:perlayer}. Each layer tensor uses a Merkle tree hashing kernel to generate the corresponding layer digest. We assign a dedicated CUDA stream to each layer, executing per-layer Merkle tree kernels concurrently. After waiting for all layer digests to be computed by synchronizing the created CUDA streams, a final Merkle tree reduction kernel computes the model digest 
The hashing scheme returns the per-layer digests and the final model digest, providing two granularity levels as discussed in Section~\ref{sec:modelauth}. Furthermore, the per-layer hashing scheme mitigates the additional memory requirements of the coalesced scheme.

\Paragraph{In-place Hashing}
The in-place hashing scheme modifies the block hashing kernel to operate on fragmented layer tensors at once as specified in Algorithm~\ref{alg:inplace}. To do so, we design a lookup table which stores, for each range of threads assigned to work on the same weight tensor, the starting address, block size, and tensor size 
for computing corresponding digests. Accordingly, when the block hashing kernel is launched, each thread within the kernel retrieves its input data by accessing the lookup table (at the row where the index range includes the thread index) and obtaining the starting address and size of input. The input size for each thread within the kernel is the same except for the threads working on the last block of every weight tensor, whereby the input data range is the starting address of the thread to the end of the tensor, which is typically less than the fixed block size. This method is used instead of simply padding the last block of data up to the block size because in-place hashing reads data directly from the state dictionary, which should be kept unchanged. After the block hashing kernel computes all the digests, the Merkle tree reduction kernel operates on the digests to obtain the final model digest. 


\subsubsection{Lattice Hashing on GPU}\label{sec:latticehash}


\sys implementation of the lattice hashing follows the high level structure of the Merkle tree implementation by incorporating a block hashing kernel and a reduction kernel. However, the main operations in each kernel of lattice hash is different from Merkle tree. The block hashing kernel incorporates Blake2b based on the GPU implementation in~\cite{mochimodevCUDAHash} for digest computation, and the reduction kernel uses the lattice addition (with GPU implementation described later) as the reduction operation in a binary tree to perform summation to reduce a group of digests into one. Because of the order invariance property of lattice hash (Section~\ref{sec:HashConstr}), we concatenate each data block with the block index before computing the hash digest for that block to mitigate reordering attacks.

\begin{algorithm}[t]
\caption{\sys In-place Hashing}\label{alg:inplace}
\begin{algorithmic}
    \State {\bfseries Input:} Model $\mathsf{state\_dict} = \{L_1:T_1,\dots,L_n:T_n\}$
    \State {\bfseries Output:} Model hash value
    \State $k\leftarrow 1$
    \For {$i=1$ to $n$}
        \State $m \leftarrow$ number of blocks in $T_i$
        \For {$j=1$ in $m$}
            \State $\mathsf{hash\_blocks}[k]\leftarrow \mathsf{BlockHash}(T_i[j])$
            \State $k \leftarrow k+1$
        \EndFor
    \EndFor
    \State $\mathsf{hash} \leftarrow \mathsf{HashReduce}(\mathsf{hash\_blocks})$
    \State \textbf{return} \textsf{hash}
\end{algorithmic}
\end{algorithm}

\begin{algorithm}[t]
\caption{\sys Optimized Lattice Per-layer}\label{alg:lattice}
\begin{algorithmic}
    \State {\bfseries Input:} Model $\mathsf{state\_dict} = \{L_1:T_1,\dots,L_n:T_n\}$
    \State {\bfseries Output:} Model hash value
    \For {$i=1$ to $n$}
        \State $m \leftarrow$ number of blocks in $T_i$
        \For {$j=1$ in $m$}
            \State $\mathsf{hash\_blocks}[i][j]\leftarrow \mathsf{BlockHash}(i\|T_i[j])$
        \EndFor
    \EndFor
    \State $\{o_1, \dots,o_n\} \leftarrow \mathsf{OrderLayerIndices}(1,\dots,n)$
    \For {$i=o_1$ to $o_n$}
        \State $\mathsf{layers\_hash}[i] \leftarrow \mathsf{HashReduce}(\mathsf{hash\_blocks[i]})$
    \EndFor
    \State $\mathsf{hash} \leftarrow \mathsf{HashReduce}(\mathsf{layers\_hash})$
    \State \textbf{return} \textsf{hash, layers\_hash}
\end{algorithmic}
\end{algorithm}

\begin{figure}[t]
\centering
    \includegraphics[width=\linewidth]{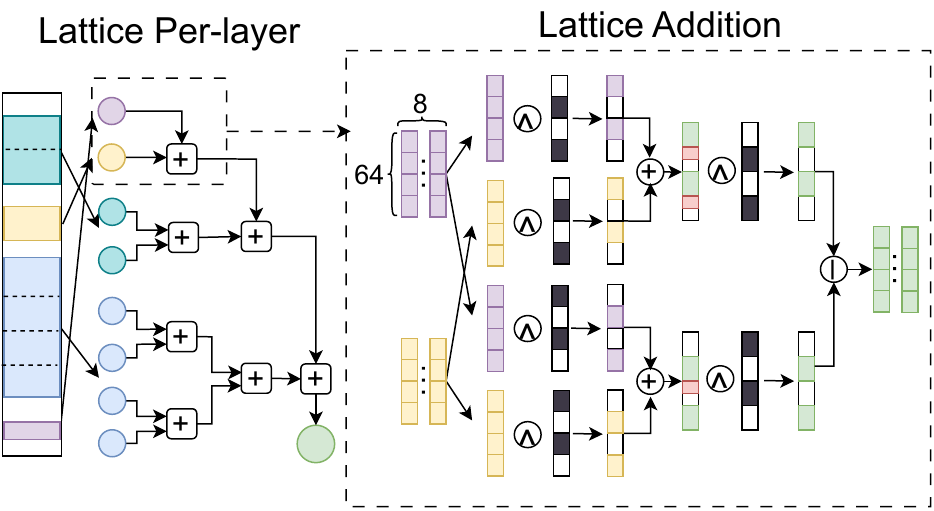}
    \caption{Details of lattice hash per-layer implementation on the left, incorporating the order-invariant property of lattice addition to sort weight tensors by size before reduction into per-layer digest and addition to running sum of model digest. The internals of lattice addition implementation for two 64-bit integers is depicted on the right.}
    \label{fig:lattice}
\end{figure}

Similar to Merkle tree implementation, lattice hashing incorporates \emph{coalesced}, \emph{per-layer}, and \emph{in-place} hashing schemes. While coalesced and in-place lattice implementations follow the structure of Merkle tree
closely, the per-layer architecture incorporates an optimization opportunity based on the order invariant property of lattice hash as shown in Figure~\ref{fig:lattice} and described in Algorithm~\ref{alg:lattice}. Specifically, after layer digests are computed, they can be accumulated into a buffer containing the running sum in any order to produce the final digest. We use this property to accumulate layer digests in ascending order of their size (represented by the \textsf{OrderLayerIndices} function in Algorithm~\ref{alg:lattice}), ensuring that smaller layer digests can be accumulated into the running sum buffer while larger layer digests are still being reduced. 
In contrast, Merkle tree per-layer hash computation requires all layer digests to be computed first, before they can be reduced to a single digest. This limitation forces smaller layers to wait for larger layer digest computation before their digest values can be reduced into the final model digest.



Incorporating the order-invariance optimization described above, the per-layer lattice hash computation starts by launching block hashing and reduction kernels for each layer tensor in ascending order of the layer size as depicted in Figure~\ref{fig:hashing}. 
The digest and reduction steps for each layer can be performed concurrently, and are scheduled into separate CUDA streams. To collect the results, the asynchronously launched kernels are synchronized with the host device in ascending order of the layer size. 

With the high level description of the lattice hashing described above, we now present the details of lattice addition and our proposed GPU architecture as depicted in Figure~\ref{fig:lattice} on the right. We follow the lattice hashing implementation in the Folly library~\cite{facebook_folly} with digest length set to 64 bytes or 8 64-bit integers. 
Each 64-bit segment of a digest is further partitioned into 4 partitions of 16 bits. 

To perform a lattice addition between two digest values (shown as purple and yellow 64$\times$8 partitioned arrays in Figure~\ref{fig:lattice}), each partition of the first digest is added to the corresponding partition of the second digest. The addition operation is modular, with the modulo set as $2^{16}$ for 16-bit partitions. In other words, when two 16-bit partitions are added together, the carry over is discarded (performing the modular operation).
The lattice addition kernel uses a group of 8 threads to operate on two digest values being added, with each thread working on a pair of 64-bit values (or 4 partitions) as depicted in Figure~\ref{fig:lattice}. Each thread performs the following operation. For each 64-bit value, the odd and even partitions are added together in parallel. This function is implemented through masking (bit-wise AND operation with zero in corresponding bits shown as black in Figure~\ref{fig:lattice}). The odd partitions are then added together and again masked to discard any overflow values shown in red, with the same operation repeated for even partitions. Finally, the odd and even partition results are combined using an OR operation to obtain the 64-bit output. 
The reduced digest from addition can be obtained by concatenating the output of 8 threads.

\subsection{Dataset Hash Computation on GPU} \label{sec:datasetHashing}

\begin{figure}
\centering
\includegraphics[width=\linewidth]{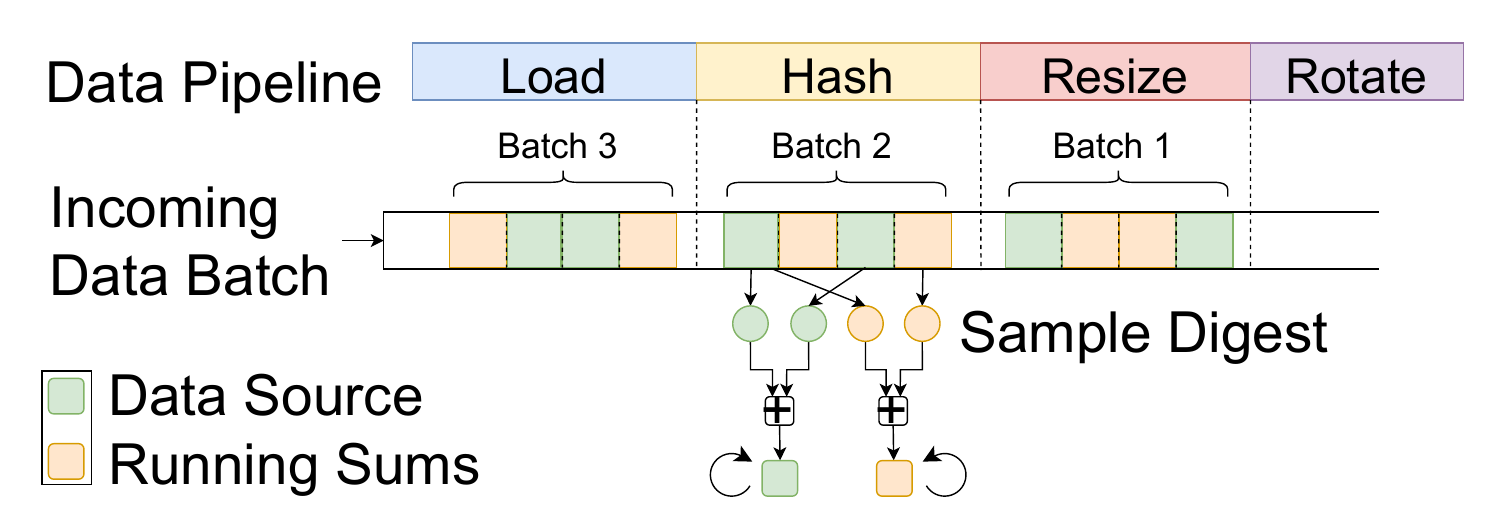}
\caption{\sys pipeline for data processing for an incoming stream of data batches. After loading each batch to GPU memory, \sys computes its lattice hash. Each batch is separated by data sources with corresponding digests accumulated into per-source digest sums. Afterwards, the data batch follows through the pipeline for other data transformations.}
\label{fig:datasetLattice}
\end{figure}

For data authentication, \sys incorporates a hashing step within the data processing pipeline on the GPU as previously detailed in Section~\ref{sec:dataauth} and shown in Figure~\ref{fig:datasetLattice}. Incoming data batches are first loaded on GPU memory, then they are processed by \sys data authenticator module which computes the data hashes (based on lattice hash), and finally they are passed through user defined transformations. The block size within the data hashing kernel is set as the size of a single sample, which are then reduced to a digest value and aggregated into a buffer holding the running sum. After all the data batches are processed, the running sum buffer holds the output digest value and is used with signature verification step. 


If a dataset is created from multiple providers, random shuffling will result in batches with data points from different sources and authenticated by separate signatures.
To address this setting, \sys implements a dictionary to map each source represented in the batch to the samples in the batch that belong to the source. With a batch size of $n$, we spawn $n$ threads for initial block hashing. Then, with $n$ digests or $\frac{n}{2}$ digest pairs to sum up, we spawn 8 threads per digest pair to reduce two digests into one. A lattice hash reduction tree is spawned for each data source represented in the incoming batch. After the trees are fully reduced, we acquire the per-source digests of the data batch. Each per-source digest is added to the running sum of digests of each source.

\subsection{\sys Library and End-to-End Operation}

\sys is implemented as a Python library for ease of use by developers and machine learning practitioners. To incorporate the model and dataset authentication components described before in a modular way, \sys uses the NVIDIA runtime compiler (NVRTC) to pre-compile the CUDA source code to callable modules from within Python during runtime.
Under the hood, \sys uses CUDA Python~\cite{CUDAPython} to call CUDA runtime and driver APIs to manage GPU memory from within the Python code base. 

We next describe an example scenario to demonstrate the end-to-end operation of \sys. Assume a user has obtained an ML model and dataset (for example retrieved from a model hub) which currently reside in their storage. 
\sys starts by loading the model to GPU memory which can be accelerated through GPUDirect Storage and bypassing the CPU. \sys performs model authentication on the fly when the model is loaded on GPU memory. \sys data pipeline incorporates user defined data transformations, and integrates data authentication as described in Section~\ref{sec:datasetHashing}. The data loader for \sys also supports reading data batches directly from storage to GPU with GPUDirect. \sys incorporates model signing as well as model and dataset signature verification by incorporating a GPU-accelerated ECDSA algorithm (based on~\cite{hkustRapidEC}) which signs or verifies the signature on each artifact authenticated payload on the GPU.





\section{Evaluation} \label{sec:Evaluation}

\begin{figure*}
\centering
\includegraphics[width=1\linewidth]{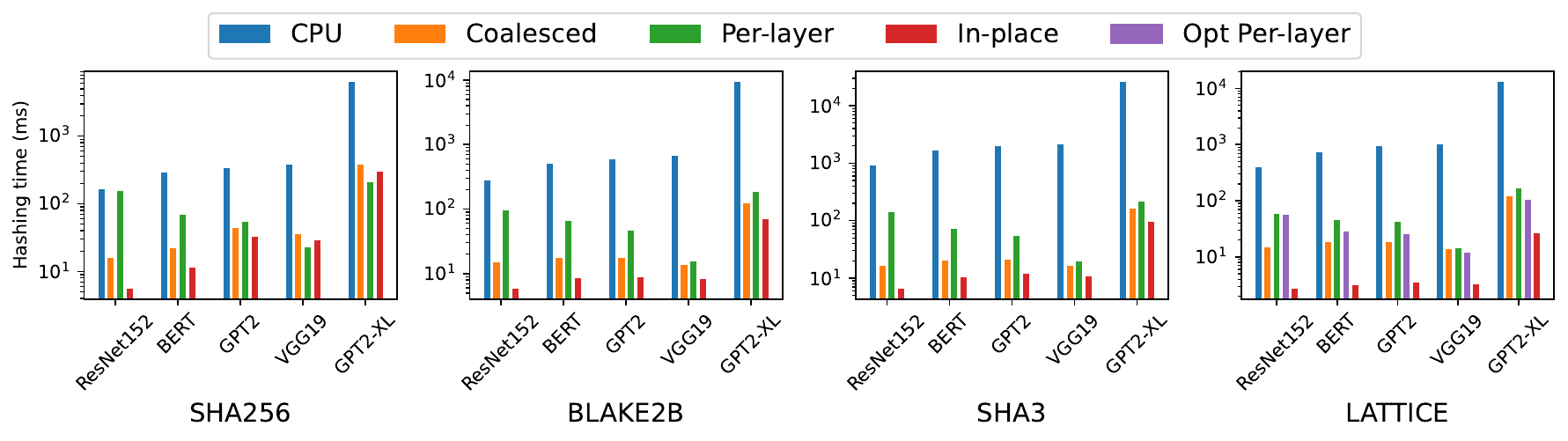}
\caption{Runtime of \sys for hashing different ML models using proposed hashing architectures (coalesced, per-layer, and in-place) compared to a CPU baseline.}
\label{fig:hash_runtime}
\end{figure*}

In this section, we quantify the performance of \sys and demonstrate its efficiency for ML artifact authentication. 

\Paragraph{Hardware Specifications} We evaluate our GPU-based library \sys and compare it against the naive Sigstore software signing on CPU.
We run all GPU experiments on an NVIDIA RTX A6000 with CUDA version 12.8 and driver version 570.86.16. The GPU has 48 GB of GDDR6 memory, 84 stream multiprocessors (SM) and a memory bandwidth of 768 GB/sec~\cite{nvidiaadaarch}. 
Our data loader implementation incorporates GPUDirect Storage in compatibility mode~\cite{cuFileAPICompat} which uses POSIX APIs to handle cuFile reads and writes through CPU memory. While GPUDirect in compatibility mode does not achieve the full speedup offered by GPUDirect, it results in file load improvements as we show later.
We run all CPU experiments on an AMD Ryzen Threadripper PRO 5995WX with 64-Cores at 4.5 GHz with 512 GB memory.

\Paragraph{Datasets and Models}
We evaluate \sys using one image dataset, namely CIFAR10~\cite{krizhevskycifar2009}) and one text dataset, namely Hellaswag~\cite{zellers2019hellaswag}. We benchmark ResNet152~\cite{he2015resnet} and VGG19~\cite{simonyan2015vgg} models for CIFAR10 and Bert~\cite{devlin2019bert}, and three variants of GPT2~\cite{radford2019language} for Hellaswag datasets.
CIFAR10 is an image data set where each sample is a 32$\times$32$\times$3 image. Hellaswag is a query data set with varying sentence lengths, using state-of-the-art generators, discriminators and source text to curate the dataset~\cite{zellers2019hellaswag}. The model specifications are summarized in Table~\ref{table:models}. 

\Paragraph{System Parameters} The default block size for hashing models is set to 8192 bytes.
For dataset hashing, the block size was set to 32$\times$32$\times$3 bytes for CIFAR10 corresponding to the size of an image, and to the size of the samples in the batch for Hellaswag where samples are padded to the same length.
The maximum number of threads per thread block is set to 512 to ensure a thread block has enough resources to use in a stream multiprocessor.

\subsection{Model Authentication Performance} \label{sec:modelruntime}
\subsubsection{Runtime Evaluation}
We start by evaluating \sys runtime for model hashing and signature verification.
Figure~\ref{fig:hash_runtime} summarizes the runtime for model digest computation of Merkle tree with SHA256, BLAKE2B, and SHA3 and comparison to Sigstore (based on hashlib~\cite{hashlib} library) baseline. Additionally, we present model digest computation runtime based on \sys lattice hash.

Among Merkle tree variations, the in-place architecture observes the lowest runtime in most settings, achieving between 11$\times$ (for GPT2 using SHA256) and 269$\times$ (for GPT2-XL using SHA3) speedup compared to the CPU baseline with Sigstore. 
The GPU runtime improvements under BLAKE2B and SHA3-based Merkle tree implementations are higher than SHA256-based implementation. This result is due to SHA256 CPU implementation benefiting from specialized hardware instruction sets. Specifically, modern Intel and AMD CPUs accelerate common cryptographic operations~\cite{intel_shaext} including SHA256 which grants a boost to CPU performance of this hash function. Nevertheless, \sys SHA256-based Merkle tree implementation achieves between 1.1$\times$ and 30$\times$ speedup compared to CPU implementation across different implementations and models.
Compared to in-place implementation, coalesced runtime is slightly higher consistently across different settings. This is because the coalesced implementation of Merkle tree incorporates an additional step which coalesces fragmented model layers in a contiguous memory block and therefore incurs additional overhead due to memory copy operations compared to in-place implementation. 
The per-layer implementation of Merkle tree typically performs worse than other implementations, since digest values for each layer has to be computed separately which limits parallelization opportunities. However, model architecture can affect concrete runtime and observed speedup as explained next.


For the largest model we benchmark, namely GPT2-XL with 1.6B parameters, we observe the greatest speedup from GPU acceleration.
ResNet152 is the smallest model we benchmark with 60M parameters, but it has the most layers at 932 as detailed in Table~\ref{table:models}. Due to the large number of layers, per-layer implementation of Merkle tree for ResNet152 incurs significantly more overhead against coalesced and in-place compared to other models as demonstrated in Figure~\ref{fig:hash_runtime}.
Although VGG19 is 2.4$\times$ larger than ResNet152 in terms of number of parameters, it has fewer number of layers (38 layers). As a result, per-layer implementation of Merkle tree when for VGG19 causes a smaller increase in runtime compared to coalesced and in-place hashing. This is evidently shown in Figure~\ref{fig:hash_runtime}, where the per-layer runtime for ResNet152 is greater than VGG19 in all hashing algorithms even though VGG19 is a larger model by parameter count.

The last plot in Figure~\ref{fig:hash_runtime} shows the performance of \sys lattice hash computation. Sigstore does not support lattice hashing, so we implement a CPU-based lattice hashing for model authentication based on Folly~\cite{facebook_folly}. As the figure shows, the comparative performance of coalesced, per-layer, and in-place implementations follows the same trends as Merkle tree hash discussed above with in-place and per-layer having the best and worst runtime respectively. Specifically, compared to the CPU baseline, \sys achieves between $5\times$ (for per-layer on ResNet152) and $500\times$ (for inplace on GPT2-XL) speedup. In this figure we also compare the performance of baseline per-layer lattice hash to our optimized per-layer lattice hash which benefits from the order invariant property as discussed in Section~\ref{sec:latticehash}. The optimized per-layer hash achieves between 5\% and 41\% speedup over the baseline per-layer implementation.


After model digests are computed, \sys creates the attestation payloads and incorporates a GPU accelerated ECDSA implementation (based on RapidEC~\cite{hkustRapidEC}) to sign the payloads. We note that the payload size depends on the number of digests inserted and is equal to 558 bytes for single digest and varies if per-layer digests are included, between 12 KB for VGG19 and 294 KB for ResNet152. Accordingly, the signing operation on GPU incurs a small overhead of 8.94 ms for single digest and varies between 10.7 ms and 38.4 ms for per-layer digests in VGG19 and ResNet152 respectively.

\begin{table}
\resizebox{\columnwidth}{!}{
\begin{tabular}{lcccccl}\toprule
& \multicolumn{3}{c}{Model Specifications} & \multicolumn{3}{c}{Additional Memory Usage}
\\\cmidrule(lr){2-4}\cmidrule(lr){5-7}
 & Weights & Layers & Size & Coalesced & Per-layer & In-place \\
 & (M) & (num) & (MB) & (MB) & (MB) & (MB) \\\midrule
ResNet152 & 60 & 932 & 270 & 270 & 2 & 4 \\
Bert & 109 & 199 & 538 & 538 & 4 & 4 \\
GPT2 & 124 & 149 & 1077 & 1077 & 8 & 6 \\
VGG19 & 143 & 38 & 1077 & 1077 & 8 & 6 \\
GPT2-XL & 1610 & 581 & 8623 & 8623 & 54 & 54 \\
\bottomrule
\end{tabular}}
\caption{Description of models and memory usage of proposed implementations over model size.}
\label{table:models}
\end{table}

\subsubsection{Memory Usage Evaluation}

Table~\ref{table:models} presents the GPU memory usage for coalesced, per-layer, and in-place implementations. The memory usage is the same for Merkle tree and lattice hash variations. We report \emph{additional} memory usage beyond the space needed to load the model. As the table shows, coalesced hashing requires the largest memory with additional allocation equal to the size of the model for coalescing model layers in a contiguous block.
Meanwhile, the memory usage for per-layer and in-place implementations are minimal since they only use buffers to store digests produced by the initial hash block kernel. Each digest is of size 32 bytes or 64 bytes depending on hashing algorithm, which is significantly smaller than the default block size of 8192 bytes. For comparison, the baseline CPU implementation incurred 2 MB of memory usage in addition to memory allocation for the ML models. 

Our results show that in-place hash implementation offers favorable performance in terms of both runtime and memory usage. However, if the underlying application requires incorporation of digest values for each layer, then per-layer implementation can be used. 

\begin{figure}
\centering
\subfloat[\sys data pipeline runtime compared to CPU baseline]{\includegraphics[width=\linewidth]{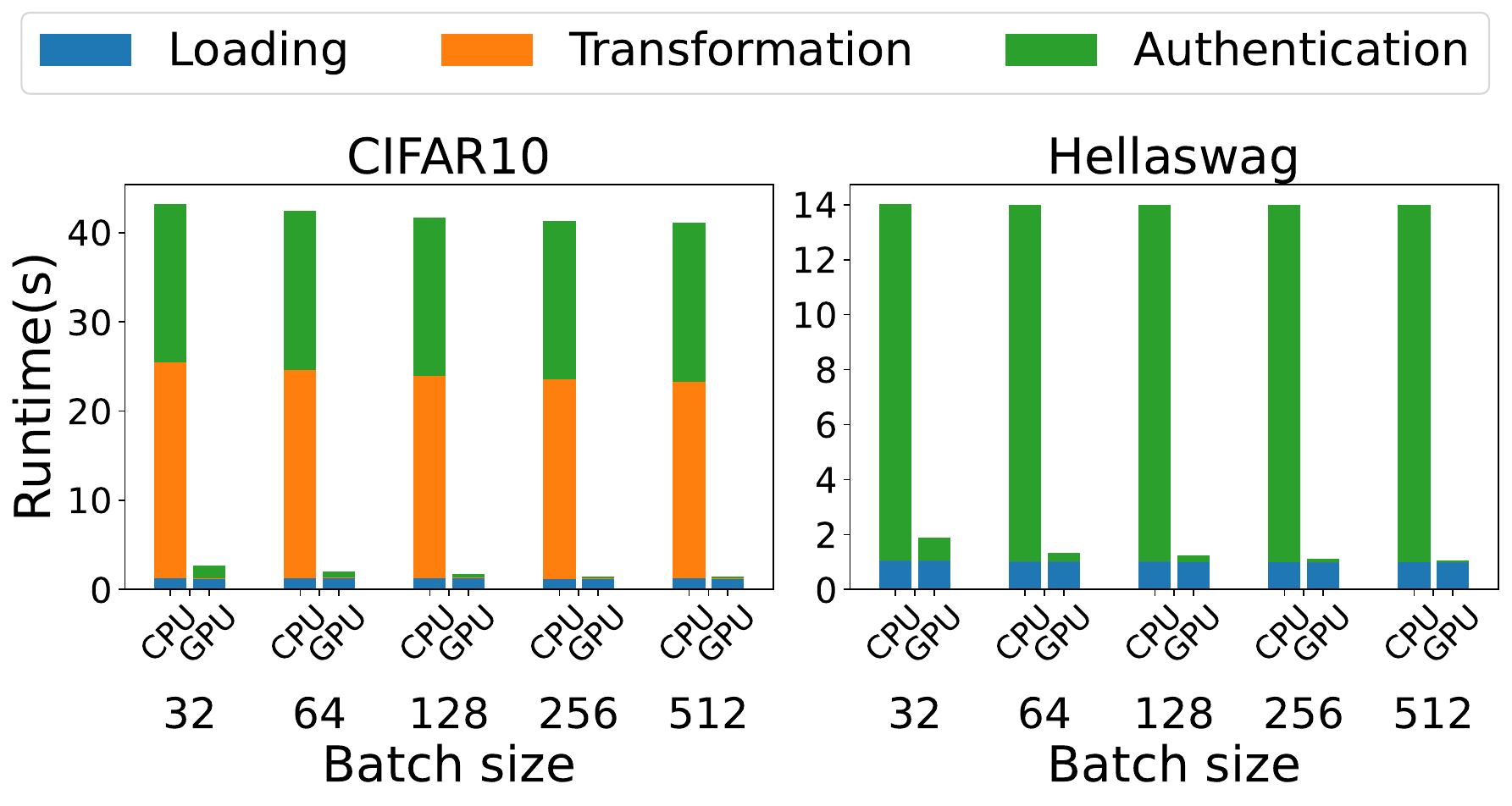}}\\
\subfloat[\sys data pipeline runtime for varying number of data sources]{\includegraphics[width=\linewidth]{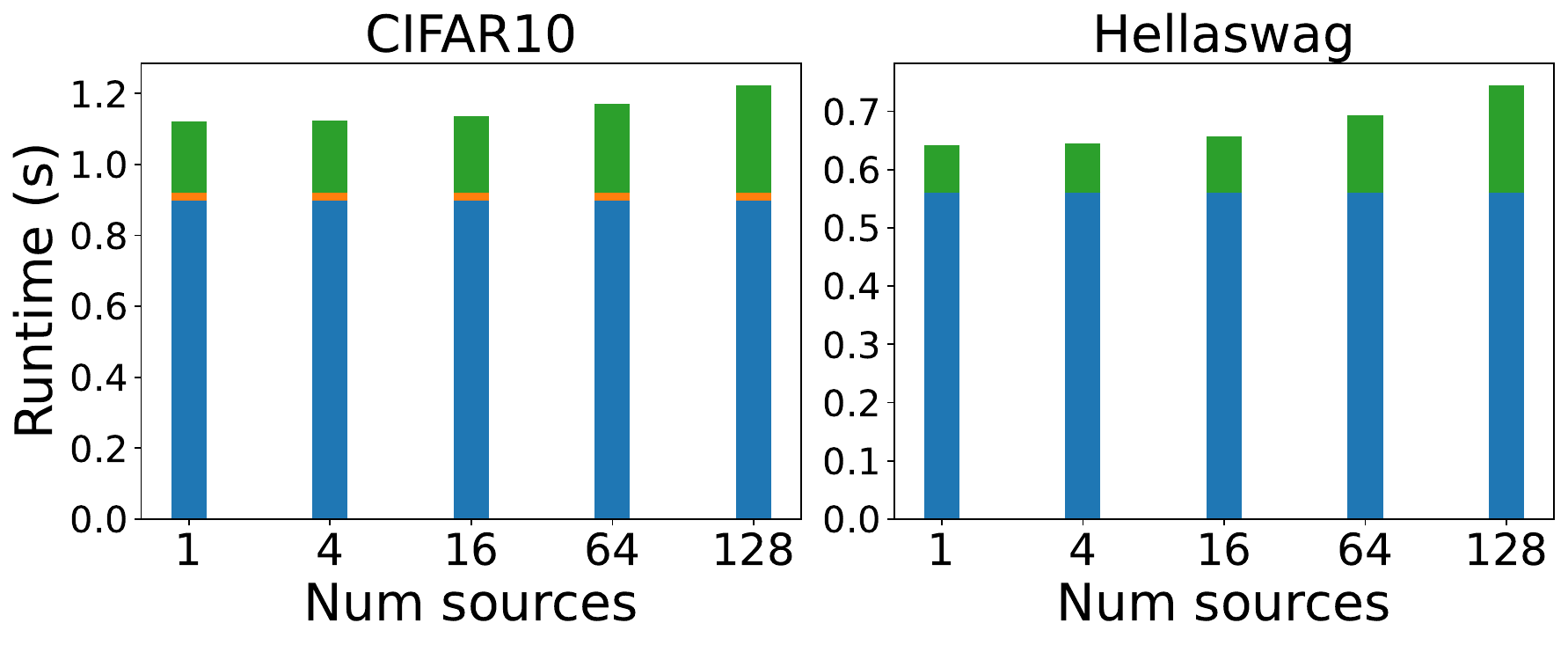}}
\caption{Runtime breakdown of \sys data pipeline, (a) comparing to CPU baseline, and (b) with varying number of data sources, for image (left) and text (right) datasets.}
\label{fig:dali_runtime}
\end{figure}


\subsection{Dataset Authentication Performance}

In this section, we evaluate the performance of \sys data pipeline. Figure~\ref{fig:dali_runtime}(a) shows the breakdown of \sys data pipeline compared to a CPU baseline (based on Sigstore) 
for CIFAR10 image and Hellaswag text datasets, assuming a batch size of 128. The data transformations implemented for CIFAR10 follow practices for processing images from NVIDIA's DALI tutorial page~\cite{nvdaligithub} and include resize, rotate, crop, mirror and normalize operations. For the text dataset, we perform padding as the sole data preprocessing step such that the inputs within a batch are of equal length. We report the runtime for processing the entire training set (50K for CIFAR and 40k for Hellaswag) for a range of batch sizes between 32 and 512 assuming 16 data sources. For \sys, we incorporate GPUDirect Storage in compatibility mode for data loading. As shown in Figure~\ref{fig:dali_runtime}(a), \sys data pipeline on GPU achieves significant speedup over CPU baseline for both datasets, resulting in up to 27$\times$ and 10$\times$ runtime improvement for CIFAR10 and Hellaswag datasets respectively.

\begin{figure*}[t]
\centering
\includegraphics[width=1\linewidth]{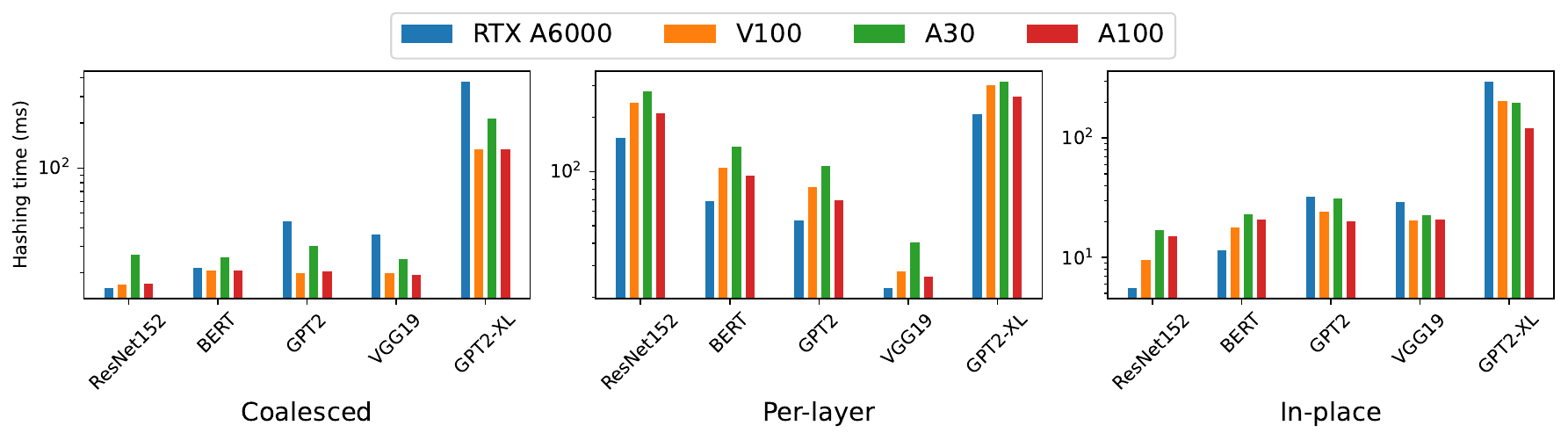}
\caption{Runtime of \sys for hashing models and datasets on various GPUs.}
\label{fig:hash_runtime_gpus}
\end{figure*}

As a percentage of total data pipeline runtime for a batch size of 32, dataset authentication including digest computation and signature verification for CIFAR10 (Hellaswag) takes up 41\% (76\%) for the CPU baseline compared to 13\% (5\%) for \sys. With a larger batch size of 512, the dataset authentication portion of total runtime falls to just 15\% (7\%). 
We note that data loading can take up to 84\% of total data processing pipeline, with 29\% runtime benefits from GPUDirect Storage running in compatibility mode. For systems that support GPUDirect Storage fully, this overhead can be further reduced.

Additionally, we evaluate the performance of \sys data pipeline with varying number of data sources as depicted in Figure~\ref{fig:dali_runtime}(b) for CIFAR10 and Hellaswag datasets. We vary the data source number from 1 to 128 and use Derichlet data partitioning~\cite{hsu2019measuringeffectsnonidenticaldata} with alpha value of 1.0 to assign data points to sources. The runtime for data processing is only impacted by the number of data sources. Therefore, while changing $\alpha$ changes the class distributions, it has no effect on total runtime. As Figure~\ref{fig:dali_runtime}(b) shows, the increase in number of data sources slightly increases the data authentication runtime (by 15\% from 1 to 128).

After dataset digest digests are computed (one per data source), \sys verifies the signatures on each payload containing the final digest. The signature verification for all data sources is executed concurrently on the GPU, incurring a small overhead of 128 ms for 128 data sources.



\Paragraph{Note on Storage Overhead} 
\sys adds attestation payloads to be stored with ML model and datasets. However, such payloads only adds a small overhead in terms of storage. To demonstrate this, assume that the payload for model contains per-layer digests and the payload for dataset contains per-sample digests with digest length of 64 Bytes. For this setting, the payload for ResNet152 (with the largest number of layers) is 294732 Bytes, adding 0.11\% overhead to storage cost.  
For datasets, the payload file contains per-sample digests and source IDs and adds only 6.5\% and 5\% storage overhead  to CIFAR10 and Hellaswag datasets respectively.


\subsection{Cross-platform Performance and Compatibility}


As previously discussed, \sys proposes new algorithms for hashing tailored for GPUs. However, the proposed algorithms are based on common constructions such as Merkle tree or SHA256 and reproducible in CPU-based systems and across different GPU architectures. While special purpose accelerators (e.g., TPUs~\cite{jouppi2017datacenter}) have to rely on CPU-based authentication, \sys ensures compatibility and orders of magnitude speedup when GPUs are available. 

In Figure~\ref{fig:hash_runtime_gpus}, we evaluate the runtime of Sentry (using Merkle hashing with SHA256) across different GPUs (RTX A6000, V100, A30, and A100) for which we confirm identical digest computations.
While there are variances in runtime across different constructions, we observe similar speedup trends with respect to the CPU runtime.
For coalesced and in-place hashing, RTX A6000 performs best on smaller models, i.e., ResNet152 and BERT, achieving up to 13.4$\times$ and 29.9$\times$ speedup compared to the CPU baseline for coalesced and in-place respectively. For larger models such as GPT2-XL, A100 achieves the best performance for coalesced and in-place hashing scheme, reaching 47.7$\times$ and 52.4$\times$ speedup respectively compared to CPU. This varying performance across models can be attributed to the fact that workstation GPUs such as RTX A6000 have more CUDA cores but lower memory bandwidth when compared to data center GPUs such as A100, enabling a greater capacity for parallel computation but slower movement of data between GPU memory regions.
Similarly, RTX A6000 performs best for per-layer hashing across all benchmarks, achieving 30$\times$ speedup over a CPU baseline, compared to A30 (least speedup) that achieves 20$\times$ runtime benefit over the CPU.

\Paragraph{Limitations:} \sys currently only provides support for NVIDIA GPUs. However, the proposed parallel algorithms and kernels can be converted to other GPU architectures which is an interesting direction for future work. For instance, HIPIFY~\cite{amdhipify} converts CUDA to portable HIP C++ code for AMD GPUs, and CUDA frameworks can be ported to SYCL for compatibility with Intel GPUs~\cite{vasquez2023sycl}.
\section{Related Work}\label{sec:related}

\Paragraph{Software Security:} Security issues in software supply chains have gained significant attention in recent years with several exiting work aiming to mitigate security risks. One such tool is Sigstore~\cite{sigstoreml} which includes features such as OIDC identity authentication, PKI for signing, and a logging feature for signatures. To protect other parts of the software supply chain beyond just software integrity, SLSA~\cite{slsa} was proposed as a set of specifications on how to perform compliant operations in every step of the supply chain to protect against supply chain attacks. As a result, SLSA allows the end user to verify the inputs, code base, and metadata that create the software. SLSA makes use of in-toto attestations to maintain the claims of performing each step correctly in a statement for later verification~\cite{slsa_intoto}.
In-toto is a framework for recording information about the steps of a software supply chain~\cite{torres2019toto}. Every party involved in the supply chain logs their actions into a payload describing the steps performed and how it was performed. Another important concept in software security is the Software Bill of Materials (SBOM) to document the list of components  that make up a software~\cite{cisa_sbom}. Nevertheless, none of the previously mentioned frameworks is specifically tailored to ML systems and scalability requirements that accompany the corresponding artifacts which we address in this work.

\Paragraph{Machine Learning Governance:} Prior work on formalizing machine learning governance include a systematization of knowledge by Chandrasekaran et al.~\cite{chandrasekaran2021sok} where they propose establishing identities in the ML supply chain through data and model ownership. The authors also highlighted the importance of model management tools to patch ML systems. In parallel with the goal of authenticating software, Google's Model Transparency~\cite{google_model_2023} initiative proposes adopting existing software authentication tools for the machine learning life cycle, incorporating the above two solutions, Sigstore and SLSA. Besides that, model card initiatives such as those incorporated by Huggingface, incorporate information about build specifications, expected results, and the checksum of each file if given by the provider. However, there is no certain way to verify their truthfulness. 

Atlas~\cite{spoczynski2025atlas} presents a framework for ML pipeline attestation and integrates the Content Provenance and Authenticity (C2PA) standard for provenance. The attestation client in Atlas runs within trusted execution environments (e.g., Intel  SGX) as a root of trust and protection against runtime tampering.
We note that \sys provides a related and orthogonal goal to the works mentioned above, and can be integrated within these frameworks to improve authentication performance. Specifically, existing work are limited to running artifact authentication on CPU resulting in significant overhead for state-of-the-art ML artifacts and incompatibility with efficient GPU data movement solutions such as GPUDirect Storage or RDMA.


\Paragraph{GPU Acceleration of Cryptographic Protocols:} Prior work has explored accelerating cryptographic primitives and protocols such as AES~\cite{manavski2007cuda, lee_efficient_2022}, hash functions~\cite{mochimodevCUDAHash}, digital signatures~\cite{hkustRapidEC}, and secure multiparty computation (MPC)~\cite{gan2025cuOT}. GPU acceleration of MPC and homomorphic encryption (HE) is of particular interest due to applications in privacy-preserving machine learning~\cite{garimella2023characterizing, anofel, tan2021cryptgpu, watson2022piranha}.
However, existing solutions do not apply to ML systems and lack specific design features to support ML system and artifact properties as discussed in this work.

\Paragraph{Confidential Computing on GPU:} Systems based on Trusted Execution Environments (TEE) and Confidential Computing (CC) on the CPU has gained popularity due to their ability to isolate memory from unauthorized sources and perform computations in isolation~\cite{sabt2015tee}.  
Recent developments of CC on GPU started with NVIDIA supporting such feature on the Hopper and Blackwell GPUs~\cite{apsey2023confidential}. CC on the GPU aims to use data isolation and restricted access from the CPU to prevent unauthorized access and tampering of data in computation. As \sys is designed to operate on the GPU with minimal CPU involvement, this feature complements the goal of \sys. We believe these recent developments in secure computation in the GPU field to be a promising prospect for an end-to-end secure AI computation on a trusted execution device.

\section{Conclusion}
This paper introduced \sys, the first GPU-based ML artifacts authentication framework that which is tailored specifically for ML systems. We presented GPU implementation of cryptographic hash constructions based on Merkle tree and lattice hashing which incorporated various GPU optimization strategies.
Additionally, we detailed \sys's custom modules for model authentication and data set authentication compatible with memory fragmentation and GPUDirect solutions.
Our benchmarks showed that \sys achieves up to 269$\times$ and 180$\times$ speedup for time critical phases of model and data set authentication respectively. Using GPUs, a compute platform that is both ubiquitous and widely used for ML training and inference, we showcased the potential of embedding security into existing ML frameworks that already run on those hardware architectures. With ever growing model sizes and data set sizes for complicated machine learning systems, and stronger demand for verifying the integrity and provenance of ML artifacts, our work opens up a new domain in GPU architectural design considerations for the future of the machine learning life cycle.

\bibliographystyle{ACM-Reference-Format}
\balance
\bibliography{mybib,refs-mlsec,refs-hwsec}


\begin{thebibliography}{69}


\ifx \showCODEN    \undefined \def \showCODEN     #1{\unskip}     \fi
\ifx \showISBNx    \undefined \def \showISBNx     #1{\unskip}     \fi
\ifx \showISBNxiii \undefined \def \showISBNxiii  #1{\unskip}     \fi
\ifx \showISSN     \undefined \def \showISSN      #1{\unskip}     \fi
\ifx \showLCCN     \undefined \def \showLCCN      #1{\unskip}     \fi
\ifx \shownote     \undefined \def \shownote      #1{#1}          \fi
\ifx \showarticletitle \undefined \def \showarticletitle #1{#1}   \fi
\ifx \showURL      \undefined \def \showURL       {\relax}        \fi
\providecommand\bibfield[2]{#2}
\providecommand\bibinfo[2]{#2}
\providecommand\natexlab[1]{#1}
\providecommand\showeprint[2][]{arXiv:#2}

\bibitem[dss({[n.\,d.]})]%
        {dsse}
 \bibinfo{year}{[n.\,d.]}\natexlab{}.
\newblock \bibinfo{title}{DSSE: Dead Simple Signing Envelope}.
\newblock
\urldef\tempurl%
\url{https://github.com/secure-systems-lab/dsse}
\showURL{%
\tempurl}


\bibitem[Almashaqbeh and Ghodsi(2025)]%
        {anofel}
\bibfield{author}{\bibinfo{person}{Ghada Almashaqbeh} {and} \bibinfo{person}{Zahra Ghodsi}.} \bibinfo{year}{2025}\natexlab{}.
\newblock \showarticletitle{ANOFEL: supporting anonymity for privacy-preserving federated learning}.
\newblock \bibinfo{journal}{\emph{Proceedings on Privacy Enhancing Technologies}} (\bibinfo{year}{2025}).
\newblock


\bibitem[AMD({[n.\,d.]})]%
        {amdhipify}
\bibfield{author}{\bibinfo{person}{AMD}.} \bibinfo{year}{[n.\,d.]}\natexlab{}.
\newblock \bibinfo{title}{HIPIFY documentation}.
\newblock
\urldef\tempurl%
\url{https://rocm.docs.amd.com/projects/HIPIFY/en/latest/}
\showURL{%
\tempurl}


\bibitem[Apsey et~al\mbox{.}(2023)]%
        {apsey2023confidential}
\bibfield{author}{\bibinfo{person}{Emily Apsey}, \bibinfo{person}{Phil Rogers}, \bibinfo{person}{Michael O'Connor}, {and} \bibinfo{person}{Rob Nertney}.} \bibinfo{year}{2023}\natexlab{}.
\newblock \bibinfo{title}{Confidential Computing on NVIDIA H100 GPUs for Secure and Trustworthy AI}.
\newblock
\urldef\tempurl%
\url{https://developer.nvidia.com/blog/confidential-computing-on-h100-gpus-for-secure-and-trustworthy-ai/}
\showURL{%
\tempurl}


\bibitem[Aumasson et~al\mbox{.}(2013)]%
        {cryptoeprint:2013/322}
\bibfield{author}{\bibinfo{person}{Jean-Philippe Aumasson}, \bibinfo{person}{Samuel Neves}, \bibinfo{person}{Zooko Wilcox-O'Hearn}, {and} \bibinfo{person}{Christian Winnerlein}.} \bibinfo{year}{2013}\natexlab{}.
\newblock \bibinfo{title}{{BLAKE2}: simpler, smaller, fast as {MD5}}.
\newblock \bibinfo{howpublished}{Cryptology {ePrint} Archive, Paper 2013/322}.
\newblock
\urldef\tempurl%
\url{https://eprint.iacr.org/2013/322}
\showURL{%
\tempurl}


\bibitem[BBC(2024)]%
        {bytedance}
\bibfield{author}{\bibinfo{person}{BBC}.} \bibinfo{year}{2024}\natexlab{}.
\newblock \bibinfo{title}{TikTok owner sacks intern for sabotaging AI project}.
\newblock
\urldef\tempurl%
\url{https://www.bbc.com/news/articles/c7v62gg49zro}
\showURL{%
\tempurl}


\bibitem[Bellare et~al\mbox{.}(1994)]%
        {bellare1994incremental}
\bibfield{author}{\bibinfo{person}{Mihir Bellare}, \bibinfo{person}{Oded Goldreich}, {and} \bibinfo{person}{Shafi Goldwasser}.} \bibinfo{year}{1994}\natexlab{}.
\newblock \showarticletitle{Incremental cryptography: The case of hashing and signing}. In \bibinfo{booktitle}{\emph{Advances in Cryptology—CRYPTO’94: 14th Annual International Cryptology Conference Santa Barbara, California, USA August 21--25, 1994 Proceedings 14}}. Springer, \bibinfo{pages}{216--233}.
\newblock


\bibitem[Bellare and Micciancio(1997)]%
        {bellare1997new}
\bibfield{author}{\bibinfo{person}{Mihir Bellare} {and} \bibinfo{person}{Daniele Micciancio}.} \bibinfo{year}{1997}\natexlab{}.
\newblock \showarticletitle{A new paradigm for collision-free hashing: Incrementality at reduced cost}. In \bibinfo{booktitle}{\emph{International Conference on the Theory and Applications of Cryptographic Techniques}}. Springer, \bibinfo{pages}{163--192}.
\newblock


\bibitem[Bertoni et~al\mbox{.}(2015)]%
        {cryptoeprint:2015/389}
\bibfield{author}{\bibinfo{person}{Guido Bertoni}, \bibinfo{person}{Joan Daemen}, \bibinfo{person}{Michael Peeters}, {and} \bibinfo{person}{Gilles~Van Assche}.} \bibinfo{year}{2015}\natexlab{}.
\newblock \bibinfo{title}{Keccak}.
\newblock \bibinfo{howpublished}{Cryptology {ePrint} Archive, Paper 2015/389}.
\newblock
\href{https://doi.org/10.1007/978-3-642-38348-9_19}{doi:\nolinkurl{10.1007/978-3-642-38348-9_19}}


\bibitem[Carlini et~al\mbox{.}(2024b)]%
        {carlini2024poisoning}
\bibfield{author}{\bibinfo{person}{Nicholas Carlini}, \bibinfo{person}{Matthew Jagielski}, \bibinfo{person}{Christopher~A Choquette-Choo}, \bibinfo{person}{Daniel Paleka}, \bibinfo{person}{Will Pearce}, \bibinfo{person}{Hyrum Anderson}, \bibinfo{person}{Andreas Terzis}, \bibinfo{person}{Kurt Thomas}, {and} \bibinfo{person}{Florian Tram{\`e}r}.} \bibinfo{year}{2024}\natexlab{b}.
\newblock \showarticletitle{Poisoning web-scale training datasets is practical}. In \bibinfo{booktitle}{\emph{2024 IEEE Symposium on Security and Privacy (SP)}}. IEEE, \bibinfo{pages}{407--425}.
\newblock


\bibitem[Carlini et~al\mbox{.}(2024a)]%
        {carlini_dataset_2024}
\bibfield{author}{\bibinfo{person}{Nicholas Carlini}, \bibinfo{person}{Matthew Jagielski}, {and} \bibinfo{person}{et al.}} \bibinfo{year}{2024}\natexlab{a}.
\newblock \showarticletitle{Poisoning Web-Scale Training Datasets is Practical}. In \bibinfo{booktitle}{\emph{2024 IEEE Symposium on Security and Privacy (SP)}}. \bibinfo{pages}{407--425}.
\newblock


\bibitem[Chandrasekaran et~al\mbox{.}(2021)]%
        {chandrasekaran2021sok}
\bibfield{author}{\bibinfo{person}{Varun Chandrasekaran}, \bibinfo{person}{Hengrui Jia}, \bibinfo{person}{Anvith Thudi}, \bibinfo{person}{Adelin Travers}, \bibinfo{person}{Mohammad Yaghini}, {and} \bibinfo{person}{Nicolas Papernot}.} \bibinfo{year}{2021}\natexlab{}.
\newblock \bibinfo{title}{SoK: Machine Learning Governance}.
\newblock
\showeprint[arxiv]{2109.10870}~[cs.CR]
\urldef\tempurl%
\url{https://arxiv.org/abs/2109.10870}
\showURL{%
\tempurl}


\bibitem[Damg{\aa}rd(1989)]%
        {damgaard1989design}
\bibfield{author}{\bibinfo{person}{Ivan~Bjerre Damg{\aa}rd}.} \bibinfo{year}{1989}\natexlab{}.
\newblock \showarticletitle{A design principle for hash functions}. In \bibinfo{booktitle}{\emph{Conference on the Theory and Application of Cryptology}}. Springer, \bibinfo{pages}{416--427}.
\newblock


\bibitem[Devlin et~al\mbox{.}(2019)]%
        {devlin2019bert}
\bibfield{author}{\bibinfo{person}{Jacob Devlin}, \bibinfo{person}{Ming-Wei Chang}, \bibinfo{person}{Kenton Lee}, {and} \bibinfo{person}{Kristina Toutanova}.} \bibinfo{year}{2019}\natexlab{}.
\newblock \bibinfo{title}{BERT: Pre-training of Deep Bidirectional Transformers for Language Understanding}.
\newblock
\showeprint[arxiv]{1810.04805}~[cs.CL]
\urldef\tempurl%
\url{https://arxiv.org/abs/1810.04805}
\showURL{%
\tempurl}


\bibitem[Face({[n.\,d.]})]%
        {hfmodels}
\bibfield{author}{\bibinfo{person}{Hugging Face}.} \bibinfo{year}{[n.\,d.]}\natexlab{}.
\newblock \bibinfo{title}{Hugging Face Models}.
\newblock
\urldef\tempurl%
\url{https://huggingface.co/models}
\showURL{%
\tempurl}


\bibitem[Facebook({[n.\,d.]})]%
        {facebook_folly}
\bibfield{author}{\bibinfo{person}{Facebook}.} \bibinfo{year}{[n.\,d.]}\natexlab{}.
\newblock \bibinfo{title}{An open-source C++ library developed and used at Facebook}.
\newblock
\urldef\tempurl%
\url{https://github.com/facebook/folly}
\showURL{%
\tempurl}


\bibitem[Feng and Tramèr(2024)]%
        {feng_backdoors_2024}
\bibfield{author}{\bibinfo{person}{Shanglun Feng} {and} \bibinfo{person}{Florian Tramèr}.} \bibinfo{year}{2024}\natexlab{}.
\newblock \showarticletitle{Privacy backdoors: stealing data with corrupted pretrained models}. In \bibinfo{booktitle}{\emph{ICML'24: Proceedings of the 41st International Conference on Machine Learning}}. \bibinfo{pages}{13326--13364}.
\newblock


\bibitem[Feng et~al\mbox{.}(2022)]%
        {hkustRapidEC}
\bibfield{author}{\bibinfo{person}{Zonghao Feng}, \bibinfo{person}{Qipeng Xie}, \bibinfo{person}{Qiong Luo}, \bibinfo{person}{Yujie Chen}, \bibinfo{person}{Haoxuan Li}, \bibinfo{person}{Huizhong Li}, {and} \bibinfo{person}{Qiang Yan}.} \bibinfo{year}{2022}\natexlab{}.
\newblock \showarticletitle{Accelerating Elliptic Curve Digital Signature Algorithms on GPUs}. In \bibinfo{booktitle}{\emph{SC22: International Conference for High Performance Computing, Networking, Storage and Analysis}}. \bibinfo{pages}{1--13}.
\newblock
\href{https://doi.org/10.1109/SC41404.2022.00032}{doi:\nolinkurl{10.1109/SC41404.2022.00032}}


\bibitem[Gan et~al\mbox{.}(2025)]%
        {gan2025cuOT}
\bibfield{author}{\bibinfo{person}{Andrew Gan}, \bibinfo{person}{Setsuna Yuki}, \bibinfo{person}{Timothy Rogers}, {and} \bibinfo{person}{Zahra Ghodsi}.} \bibinfo{year}{2025}\natexlab{}.
\newblock \bibinfo{title}{cuOT: Accelerating Oblivious Transfer on GPUs for Privacy-preserving Computation}.
\newblock \bibinfo{howpublished}{IEEE International Symposium on Hardware Oriented Security and Trust (HOST)}.
\newblock


\bibitem[Garimella et~al\mbox{.}(2023)]%
        {garimella2023characterizing}
\bibfield{author}{\bibinfo{person}{Karthik Garimella}, \bibinfo{person}{Zahra Ghodsi}, \bibinfo{person}{Nandan~Kumar Jha}, \bibinfo{person}{Siddharth Garg}, {and} \bibinfo{person}{Brandon Reagen}.} \bibinfo{year}{2023}\natexlab{}.
\newblock \showarticletitle{Characterizing and optimizing end-to-end systems for private inference}. In \bibinfo{booktitle}{\emph{Proceedings of the 28th ACM International Conference on Architectural Support for Programming Languages and Operating Systems, Volume 3}}. \bibinfo{pages}{89--104}.
\newblock


\bibitem[Geer et~al\mbox{.}(2020)]%
        {geer_quant_2020}
\bibfield{author}{\bibinfo{person}{Dan Geer}, \bibinfo{person}{Bentz Tozer}, {and} \bibinfo{person}{John~Speed Meyers}.} \bibinfo{year}{2020}\natexlab{}.
\newblock \showarticletitle{For Good Measure Counting Broken Links: A Quant’s View of Software Supply Chain Security}.
\newblock \bibinfo{journal}{\emph{;login:}} \bibinfo{volume}{45}, \bibinfo{number}{4} (\bibinfo{year}{2020}), \bibinfo{pages}{83--86}.
\newblock


\bibitem[Grosse et~al\mbox{.}(2024)]%
        {grosse2024your}
\bibfield{author}{\bibinfo{person}{Kathrin Grosse}, \bibinfo{person}{Lukas Bieringer}, \bibinfo{person}{Tarek~R Besold}, \bibinfo{person}{Battista Biggio}, {and} \bibinfo{person}{Alexandre Alahi}.} \bibinfo{year}{2024}\natexlab{}.
\newblock \showarticletitle{When your AI becomes a target: AI security incidents and best practices}. In \bibinfo{booktitle}{\emph{Proceedings of the AAAI Conference on Artificial Intelligence}}, Vol.~\bibinfo{volume}{38}. \bibinfo{pages}{23041--23046}.
\newblock


\bibitem[Gu et~al\mbox{.}(2019a)]%
        {gu_badnets_2017}
\bibfield{author}{\bibinfo{person}{Tianyu Gu}, \bibinfo{person}{Brendan Dolan-Gavitt}, {and} \bibinfo{person}{Siddharth Garg}.} \bibinfo{year}{2019}\natexlab{a}.
\newblock \bibinfo{title}{BadNets: Identifying Vulnerabilities in the Machine Learning Model Supply Chain}.
\newblock
\showeprint[arxiv]{1708.06733}~[cs.CR]
\urldef\tempurl%
\url{https://arxiv.org/abs/1708.06733}
\showURL{%
\tempurl}


\bibitem[Gu et~al\mbox{.}(2019b)]%
        {gu2019badnets}
\bibfield{author}{\bibinfo{person}{Tianyu Gu}, \bibinfo{person}{Kang Liu}, \bibinfo{person}{Brendan Dolan-Gavitt}, {and} \bibinfo{person}{Siddharth Garg}.} \bibinfo{year}{2019}\natexlab{b}.
\newblock \showarticletitle{Badnets: Evaluating backdooring attacks on deep neural networks}.
\newblock \bibinfo{journal}{\emph{IEEE Access}}  \bibinfo{volume}{7} (\bibinfo{year}{2019}), \bibinfo{pages}{47230--47244}.
\newblock


\bibitem[Hashlib({[n.\,d.]})]%
        {hashlib}
Hashlib \bibinfo{year}{[n.\,d.]}\natexlab{}.
\newblock \bibinfo{title}{hashlib — Secure hashes and message digests}.
\newblock
\urldef\tempurl%
\url{https://docs.python.org/3/library/hashlib.html}
\showURL{%
\tempurl}


\bibitem[He et~al\mbox{.}(2015)]%
        {he2015resnet}
\bibfield{author}{\bibinfo{person}{Kaiming He}, \bibinfo{person}{Xiangyu Zhang}, \bibinfo{person}{Shaoqing Ren}, {and} \bibinfo{person}{Jian Sun}.} \bibinfo{year}{2015}\natexlab{}.
\newblock \bibinfo{title}{Deep Residual Learning for Image Recognition}.
\newblock
\showeprint[arxiv]{1512.03385}~[cs.CV]
\urldef\tempurl%
\url{https://arxiv.org/abs/1512.03385}
\showURL{%
\tempurl}


\bibitem[Hsu et~al\mbox{.}(2019)]%
        {hsu2019measuringeffectsnonidenticaldata}
\bibfield{author}{\bibinfo{person}{Tzu-Ming~Harry Hsu}, \bibinfo{person}{Hang Qi}, {and} \bibinfo{person}{Matthew Brown}.} \bibinfo{year}{2019}\natexlab{}.
\newblock \bibinfo{title}{Measuring the Effects of Non-Identical Data Distribution for Federated Visual Classification}.
\newblock
\showeprint[arxiv]{1909.06335}~[cs.LG]
\urldef\tempurl%
\url{https://arxiv.org/abs/1909.06335}
\showURL{%
\tempurl}


\bibitem[in~toto({[n.\,d.]})]%
        {intotospec}
\bibfield{author}{\bibinfo{person}{in toto}.} \bibinfo{year}{[n.\,d.]}\natexlab{}.
\newblock \bibinfo{title}{in-toto Specification}.
\newblock
\urldef\tempurl%
\url{https://github.com/in-toto/specification/blob/v1.0/in-toto-spec.md}
\showURL{%
\tempurl}


\bibitem[Intel({[n.\,d.]})]%
        {intel_shaext}
\bibfield{author}{\bibinfo{person}{Intel}.} \bibinfo{year}{[n.\,d.]}\natexlab{}.
\newblock \bibinfo{title}{New Instructions Supporting the Secure Hash Algorithm on Intel® Architecture Processors}.
\newblock
\urldef\tempurl%
\url{https://www.intel.com/content/www/us/en/developer/articles/technical/intel-sha-extensions.html}
\showURL{%
\tempurl}


\bibitem[Jouppi et~al\mbox{.}(2017)]%
        {jouppi2017datacenter}
\bibfield{author}{\bibinfo{person}{Norman~P Jouppi}, \bibinfo{person}{Cliff Young}, \bibinfo{person}{Nishant Patil}, \bibinfo{person}{David Patterson}, \bibinfo{person}{Gaurav Agrawal}, \bibinfo{person}{Raminder Bajwa}, \bibinfo{person}{Sarah Bates}, \bibinfo{person}{Suresh Bhatia}, \bibinfo{person}{Nan Boden}, \bibinfo{person}{Al Borchers}, {et~al\mbox{.}}} \bibinfo{year}{2017}\natexlab{}.
\newblock \showarticletitle{In-datacenter performance analysis of a tensor processing unit}. In \bibinfo{booktitle}{\emph{Proceedings of the 44th annual international symposium on computer architecture}}. \bibinfo{pages}{1--12}.
\newblock


\bibitem[Kaggle({[n.\,d.]})]%
        {kmodels}
\bibfield{author}{\bibinfo{person}{Kaggle}.} \bibinfo{year}{[n.\,d.]}\natexlab{}.
\newblock \bibinfo{title}{Kaggle Models}.
\newblock
\urldef\tempurl%
\url{https://www.kaggle.com/models}
\showURL{%
\tempurl}


\bibitem[Krizhevsky({[n.\,d.]})]%
        {krizhevskycifar2009}
\bibfield{author}{\bibinfo{person}{Alex Krizhevsky}.} \bibinfo{year}{[n.\,d.]}\natexlab{}.
\newblock \bibinfo{title}{Learning Multiple Layers of Features from Tiny Images}.
\newblock
\urldef\tempurl%
\url{https://www.cs.toronto.edu/~kriz/learning-features-2009-TR.pdf}
\showURL{%
\tempurl}


\bibitem[Lee et~al\mbox{.}(2022)]%
        {lee_efficient_2022}
\bibfield{author}{\bibinfo{person}{Wai-Kong Lee}, \bibinfo{person}{Hwa~Jeong Seo}, \bibinfo{person}{Seog~Chung Seo}, {and} \bibinfo{person}{Seong~Oun Hwang}.} \bibinfo{year}{2022}\natexlab{}.
\newblock \showarticletitle{Efficient {Implementation} of {AES}-{CTR} and {AES}-{ECB} on {GPUs} {With} {Applications} for {High}-{Speed} {FrodoKEM} and {Exhaustive} {Key} {Search}}.
\newblock \bibinfo{journal}{\emph{IEEE Transactions on Circuits and Systems II: Express Briefs}} (\bibinfo{date}{June} \bibinfo{year}{2022}), \bibinfo{pages}{2962--2966}.
\newblock


\bibitem[Lewi et~al\mbox{.}(2019)]%
        {lewi2019securing}
\bibfield{author}{\bibinfo{person}{Kevin Lewi}, \bibinfo{person}{Wonho Kim}, \bibinfo{person}{Ilya Maykov}, {and} \bibinfo{person}{Stephen Weis}.} \bibinfo{year}{2019}\natexlab{}.
\newblock \showarticletitle{Securing update propagation with homomorphic hashing}.
\newblock \bibinfo{journal}{\emph{Cryptology ePrint Archive}} (\bibinfo{year}{2019}).
\newblock


\bibitem[Manavski(2007)]%
        {manavski2007cuda}
\bibfield{author}{\bibinfo{person}{Svetlin~A Manavski}.} \bibinfo{year}{2007}\natexlab{}.
\newblock \showarticletitle{CUDA compatible GPU as an efficient hardware accelerator for AES cryptography}. In \bibinfo{booktitle}{\emph{2007 IEEE International Conference on Signal Processing and Communications}}. IEEE, \bibinfo{pages}{65--68}.
\newblock


\bibitem[Maruseac et~al\mbox{.}({[n.\,d.]})]%
        {google_model_2023}
\bibfield{author}{\bibinfo{person}{Mihai Maruseac}, \bibinfo{person}{Sarah Meiklejohn}, \bibinfo{person}{Mark Lodato}, {and} \bibinfo{person}{Google Open Source Security~Team (GOSST)}.} \bibinfo{year}{[n.\,d.]}\natexlab{}.
\newblock \bibinfo{title}{Increasing transparency in AI security}.
\newblock
\urldef\tempurl%
\url{https://security.googleblog.com/2023/10/increasing-transparency-in-ai-security.html}
\showURL{%
\tempurl}
\newblock
\shownote{Accessed: 2024-09-27}.


\bibitem[Merkle(1979)]%
        {merkle1979secrecy}
\bibfield{author}{\bibinfo{person}{Ralph~Charles Merkle}.} \bibinfo{year}{1979}\natexlab{}.
\newblock \bibinfo{booktitle}{\emph{Secrecy, authentication, and public key systems.}}
\newblock \bibinfo{publisher}{Stanford university}.
\newblock


\bibitem[Merkle(1989)]%
        {merkle1989one}
\bibfield{author}{\bibinfo{person}{Ralph~C Merkle}.} \bibinfo{year}{1989}\natexlab{}.
\newblock \showarticletitle{One way hash functions and DES}. In \bibinfo{booktitle}{\emph{Conference on the Theory and Application of Cryptology}}. Springer, \bibinfo{pages}{428--446}.
\newblock


\bibitem[ModelScope({[n.\,d.]})]%
        {modelscope}
\bibfield{author}{\bibinfo{person}{ModelScope}.} \bibinfo{year}{[n.\,d.]}\natexlab{}.
\newblock \bibinfo{title}{ModelScope Models}.
\newblock
\urldef\tempurl%
\url{https://modelscope.cn/models}
\showURL{%
\tempurl}


\bibitem[Narayanan et~al\mbox{.}(2021)]%
        {narayanan2021efficient}
\bibfield{author}{\bibinfo{person}{Deepak Narayanan}, \bibinfo{person}{Mohammad Shoeybi}, \bibinfo{person}{Jared Casper}, \bibinfo{person}{Patrick LeGresley}, \bibinfo{person}{Mostofa Patwary}, \bibinfo{person}{Vijay Korthikanti}, \bibinfo{person}{Dmitri Vainbrand}, \bibinfo{person}{Prethvi Kashinkunti}, \bibinfo{person}{Julie Bernauer}, \bibinfo{person}{Bryan Catanzaro}, {et~al\mbox{.}}} \bibinfo{year}{2021}\natexlab{}.
\newblock \showarticletitle{Efficient large-scale language model training on gpu clusters using megatron-lm}. In \bibinfo{booktitle}{\emph{Proceedings of the international conference for high performance computing, networking, storage and analysis}}. \bibinfo{pages}{1--15}.
\newblock


\bibitem[Newman et~al\mbox{.}(2022)]%
        {newman2022sigstore}
\bibfield{author}{\bibinfo{person}{Zachary Newman}, \bibinfo{person}{John~Speed Meyers}, {and} \bibinfo{person}{Santiago Torres-Arias}.} \bibinfo{year}{2022}\natexlab{}.
\newblock \showarticletitle{Sigstore: Software signing for everybody}. In \bibinfo{booktitle}{\emph{Proceedings of the 2022 ACM SIGSAC Conference on Computer and Communications Security}}. \bibinfo{pages}{2353--2367}.
\newblock


\bibitem[NIST({[n.\,d.]})]%
        {nist_sha256}
\bibfield{author}{\bibinfo{person}{NIST}.} \bibinfo{year}{[n.\,d.]}\natexlab{}.
\newblock \bibinfo{title}{Secure Hash Standard (SHS)}.
\newblock
\urldef\tempurl%
\url{https://nvlpubs.nist.gov/nistpubs/FIPS/NIST.FIPS.180-4.pdf}
\showURL{%
\tempurl}


\bibitem[NVIDIA({[n.\,d.]})]%
        {CUDAPython}
\bibfield{author}{\bibinfo{person}{NVIDIA}.} \bibinfo{year}{[n.\,d.]}\natexlab{}.
\newblock \bibinfo{title}{CUDA Python 12.8.0 documentation}.
\newblock
\urldef\tempurl%
\url{https://nvidia.github.io/cuda-python/latest/}
\showURL{%
\tempurl}


\bibitem[Nvidia({[n.\,d.]})]%
        {gpudirect}
\bibfield{author}{\bibinfo{person}{Nvidia}.} \bibinfo{year}{[n.\,d.]}\natexlab{}.
\newblock \bibinfo{title}{NVIDIA Magnum IO GPUDirect Storage Overview Guide}.
\newblock
\urldef\tempurl%
\url{https://docs.nvidia.com/gpudirect-storage/overview-guide/index.html}
\showURL{%
\tempurl}


\bibitem[NVIDIA ADA({[n.\,d.]})]%
        {nvidiaadaarch}
NVIDIA ADA \bibinfo{year}{[n.\,d.]}\natexlab{}.
\newblock \bibinfo{title}{NVIDIA ADA LOVELACE PROFESSIONAL GPU ARCHITECTURE}.
\newblock
\urldef\tempurl%
\url{https://images.nvidia.com/aem-dam/en-zz/Solutions/technologies/NVIDIA-ADA-GPU-PROVIZ-Architecture-Whitepaper\_1.1.pdf}
\showURL{%
\tempurl}


\bibitem[NVIDIA CUDA({[n.\,d.]})]%
        {cudaguide}
NVIDIA CUDA \bibinfo{year}{[n.\,d.]}\natexlab{}.
\newblock \bibinfo{title}{CUDA C++ Programming Guide}.
\newblock
\urldef\tempurl%
\url{https://docs.nvidia.com/cuda/cuda-c-programming-guide/index.html}
\showURL{%
\tempurl}


\bibitem[NVIDIA cuFile API Reference({[n.\,d.]})]%
        {cuFileAPICompat}
NVIDIA cuFile API Reference \bibinfo{year}{[n.\,d.]}\natexlab{}.
\newblock \bibinfo{title}{3.3. cuFile Compatibility Mode}.
\newblock
\urldef\tempurl%
\url{https://docs.nvidia.com/gpudirect-storage/api-reference-guide/index.html#cufile-compatibility-mode}
\showURL{%
\tempurl}


\bibitem[NVIDIA DALI({[n.\,d.]})]%
        {nvdali}
NVIDIA DALI \bibinfo{year}{[n.\,d.]}\natexlab{}.
\newblock \bibinfo{title}{NVIDIA Data Loading Library}.
\newblock
\urldef\tempurl%
\url{https://docs.nvidia.com/deeplearning/dali/user\-guide/docs/index.html}
\showURL{%
\tempurl}


\bibitem[NVIDIA DALI GitHub({[n.\,d.]})]%
        {nvdaligithub}
NVIDIA DALI GitHub \bibinfo{year}{[n.\,d.]}\natexlab{}.
\newblock \bibinfo{title}{NVIDIA DALI GitHub}.
\newblock
\urldef\tempurl%
\url{https://github.com/NVIDIA/DALI}
\showURL{%
\tempurl}


\bibitem[ONNX({[n.\,d.]})]%
        {onnxmodels}
\bibfield{author}{\bibinfo{person}{ONNX}.} \bibinfo{year}{[n.\,d.]}\natexlab{}.
\newblock \bibinfo{title}{ONNX Model Zoo}.
\newblock
\urldef\tempurl%
\url{https://onnx.ai/models/}
\showURL{%
\tempurl}


\bibitem[Radford et~al\mbox{.}(2019)]%
        {radford2019language}
\bibfield{author}{\bibinfo{person}{Alec Radford}, \bibinfo{person}{Jeffrey Wu}, \bibinfo{person}{Rewon Child}, \bibinfo{person}{David Luan}, \bibinfo{person}{Dario Amodei}, {and} \bibinfo{person}{Ilya Sutskever}.} \bibinfo{year}{2019}\natexlab{}.
\newblock \showarticletitle{Language Models are Unsupervised Multitask Learners}.
\newblock \bibinfo{journal}{\emph{OpenAI}} (\bibinfo{year}{2019}).
\newblock
\urldef\tempurl%
\url{https://cdn.openai.com/better-language-models/language_models_are_unsupervised_multitask_learners.pdf}
\showURL{%
\tempurl}
\newblock
\shownote{Accessed: 2024-11-15}.


\bibitem[Reichert and Obelheiro(2024)]%
        {reichert2024software}
\bibfield{author}{\bibinfo{person}{Beatriz~M Reichert} {and} \bibinfo{person}{Rafael~R Obelheiro}.} \bibinfo{year}{2024}\natexlab{}.
\newblock \showarticletitle{Software supply chain security: a systematic literature review}.
\newblock \bibinfo{journal}{\emph{International Journal of Computers and Applications}} \bibinfo{volume}{46}, \bibinfo{number}{10} (\bibinfo{year}{2024}), \bibinfo{pages}{853--867}.
\newblock


\bibitem[Rogaway and Shrimpton(2004)]%
        {rogaway2004chf}
\bibfield{author}{\bibinfo{person}{Phillip Rogaway} {and} \bibinfo{person}{Thomas Shrimpton}.} \bibinfo{year}{2004}\natexlab{}.
\newblock \bibinfo{title}{Cryptographic Hash-Function Basics: Definitions, Implications, and Separations for Preimage Resistance, Second-Preimage Resistance, and Collision Resistance}.
\newblock
\urldef\tempurl%
\url{https://www.iacr.org/archive/fse2004/30170373/30170373.pdf}
\showURL{%
\tempurl}


\bibitem[Sabt et~al\mbox{.}(2015)]%
        {sabt2015tee}
\bibfield{author}{\bibinfo{person}{Mohamed Sabt}, \bibinfo{person}{Mohammed Achemlal}, {and} \bibinfo{person}{Abdelmadjid Bouabdallah}.} \bibinfo{year}{2015}\natexlab{}.
\newblock \showarticletitle{Trusted Execution Environment: What It is, and What It is Not}. \bibinfo{pages}{57--64}.
\newblock
\href{https://doi.org/10.1109/Trustcom.2015.357}{doi:\nolinkurl{10.1109/Trustcom.2015.357}}


\bibitem[SBOM FAQ({[n.\,d.]})]%
        {cisa_sbom}
SBOM FAQ \bibinfo{year}{[n.\,d.]}\natexlab{}.
\newblock
\urldef\tempurl%
\url{https://www.cisa.gov/sites/default/files/2024-07/SBOM%20FAQ%202024.pdf}
\showURL{%
\tempurl}


\bibitem[Sigstore({[n.\,d.]})]%
        {sigstoreml}
\bibfield{author}{\bibinfo{person}{Sigstore}.} \bibinfo{year}{[n.\,d.]}\natexlab{}.
\newblock \bibinfo{title}{Model Transparency}.
\newblock
\urldef\tempurl%
\url{https://github.com/sigstore/model-transparency}
\showURL{%
\tempurl}


\bibitem[Simonyan and Zisserman(2015)]%
        {simonyan2015vgg}
\bibfield{author}{\bibinfo{person}{Karen Simonyan} {and} \bibinfo{person}{Andrew Zisserman}.} \bibinfo{year}{2015}\natexlab{}.
\newblock \bibinfo{title}{Very Deep Convolutional Networks for Large-Scale Image Recognition}.
\newblock
\showeprint[arxiv]{1409.1556}~[cs.CV]
\urldef\tempurl%
\url{https://arxiv.org/abs/1409.1556}
\showURL{%
\tempurl}


\bibitem[SLSA({[n.\,d.]})]%
        {slsa}
SLSA \bibinfo{year}{[n.\,d.]}\natexlab{}.
\newblock
\urldef\tempurl%
\url{https://slsa.dev/}
\showURL{%
\tempurl}


\bibitem[SLSA In-toto({[n.\,d.]})]%
        {slsa_intoto}
SLSA In-toto \bibinfo{year}{[n.\,d.]}\natexlab{}.
\newblock
\urldef\tempurl%
\url{https://slsa.dev/blog/2023/05/in-toto-and-slsa}
\showURL{%
\tempurl}


\bibitem[Solis-Vasquez et~al\mbox{.}(2023)]%
        {vasquez2023sycl}
\bibfield{author}{\bibinfo{person}{Leonardo Solis-Vasquez}, \bibinfo{person}{Edward Mascarenhas}, {and} \bibinfo{person}{Andreas Koch}.} \bibinfo{year}{2023}\natexlab{}.
\newblock \showarticletitle{Experiences Migrating CUDA to SYCL: A Molecular Docking Case Study}. In \bibinfo{booktitle}{\emph{Proceedings of the 2023 International Workshop on OpenCL}}.
\newblock


\bibitem[Spoczynski et~al\mbox{.}(2025)]%
        {spoczynski2025atlas}
\bibfield{author}{\bibinfo{person}{Marcin Spoczynski}, \bibinfo{person}{Marcela~S. Melara}, {and} \bibinfo{person}{Sebastian Szyller}.} \bibinfo{year}{2025}\natexlab{}.
\newblock \bibinfo{title}{Atlas: A Framework for ML Lifecycle Provenance \& Transparency}.
\newblock
\showeprint[arxiv]{2502.19567}~[cs.CR]
\urldef\tempurl%
\url{https://arxiv.org/abs/2502.19567}
\showURL{%
\tempurl}


\bibitem[Springer(2001)]%
        {johnson2001ecdsa}
Springer \bibinfo{year}{2001}\natexlab{}.
\newblock \bibinfo{booktitle}{\emph{The Elliptic Curve Digital Signature Algorithm (ECDSA)}}. Springer.
\newblock
\href{https://doi.org/10.1007/s102070100002}{doi:\nolinkurl{10.1007/s102070100002}}


\bibitem[Tan et~al\mbox{.}(2021)]%
        {tan2021cryptgpu}
\bibfield{author}{\bibinfo{person}{Sijun Tan}, \bibinfo{person}{Brian Knott}, \bibinfo{person}{Yuan Tian}, {and} \bibinfo{person}{David~J Wu}.} \bibinfo{year}{2021}\natexlab{}.
\newblock \showarticletitle{CryptGPU: Fast privacy-preserving machine learning on the GPU}. In \bibinfo{booktitle}{\emph{2021 IEEE Symposium on Security and Privacy (SP)}}. IEEE, \bibinfo{pages}{1021--1038}.
\newblock


\bibitem[Torres-Arias et~al\mbox{.}(2019)]%
        {torres2019toto}
\bibfield{author}{\bibinfo{person}{Santiago Torres-Arias}, \bibinfo{person}{Hammad Afzali}, \bibinfo{person}{Trishank~Karthik Kuppusamy}, \bibinfo{person}{Reza Curtmola}, {and} \bibinfo{person}{Justin Cappos}.} \bibinfo{year}{2019}\natexlab{}.
\newblock \showarticletitle{in-toto: Providing farm-to-table guarantees for bits and bytes}. In \bibinfo{booktitle}{\emph{28th USENIX Security Symposium (USENIX Security 19)}}. \bibinfo{pages}{1393--1410}.
\newblock


\bibitem[Volos et~al\mbox{.}(2018)]%
        {volos2018graviton}
\bibfield{author}{\bibinfo{person}{Stavros Volos}, \bibinfo{person}{Kapil Vaswani}, {and} \bibinfo{person}{Rodrigo Bruno}.} \bibinfo{year}{2018}\natexlab{}.
\newblock \showarticletitle{Graviton: trusted execution environments on GPUs}. In \bibinfo{booktitle}{\emph{Proceedings of the 13th USENIX Conference on Operating Systems Design and Implementation}}. \bibinfo{publisher}{USENIX Association}.
\newblock


\bibitem[Watson et~al\mbox{.}(2022)]%
        {watson2022piranha}
\bibfield{author}{\bibinfo{person}{Jean-Luc Watson}, \bibinfo{person}{Sameer Wagh}, {and} \bibinfo{person}{Raluca~Ada Popa}.} \bibinfo{year}{2022}\natexlab{}.
\newblock \showarticletitle{Piranha: A $\{$GPU$\}$ platform for secure computation}. In \bibinfo{booktitle}{\emph{31st USENIX Security Symposium (USENIX Security 22)}}. \bibinfo{pages}{827--844}.
\newblock


\bibitem[Yudha et~al\mbox{.}(2022)]%
        {yudha2022lite}
\bibfield{author}{\bibinfo{person}{Ardhi Wiratama~Baskara Yudha}, \bibinfo{person}{Jake Meyer}, \bibinfo{person}{Shougang Yuan}, \bibinfo{person}{Huiyang Zhou}, {and} \bibinfo{person}{Yan Solihin}.} \bibinfo{year}{2022}\natexlab{}.
\newblock \showarticletitle{LITE: a low-cost practical inter-operable GPU TEE}. In \bibinfo{booktitle}{\emph{Proceedings of the 36th ACM International Conference on Supercomputing}}. \bibinfo{publisher}{Association for Computing Machinery}.
\newblock


\bibitem[Zellers et~al\mbox{.}(2019)]%
        {zellers2019hellaswag}
\bibfield{author}{\bibinfo{person}{Rowan Zellers}, \bibinfo{person}{Ari Holtzman}, \bibinfo{person}{Yonatan Bisk}, \bibinfo{person}{Ali Farhadi}, {and} \bibinfo{person}{Yejin Choi}.} \bibinfo{year}{2019}\natexlab{}.
\newblock \showarticletitle{HellaSwag: Can a Machine Really Finish Your Sentence?}. In \bibinfo{booktitle}{\emph{Proceedings of the 57th Annual Meeting of the Association for Computational Linguistics}}.
\newblock


\bibitem[Zweil and Team({[n.\,d.]})]%
        {mochimodevCUDAHash}
\bibfield{author}{\bibinfo{person}{Matt Zweil} {and} \bibinfo{person}{The Mochimo Core~Contributor Team}.} \bibinfo{year}{[n.\,d.]}\natexlab{}.
\newblock \bibinfo{title}{CUDA Hashing Algorithms Collection}.
\newblock
\urldef\tempurl%
\url{https://github.com/mochimodev/cuda-hashing-algos}
\showURL{%
\tempurl}


\end{thebibliography}

\end{document}